%% file: AAAI-LiuZ.9223.tex
\def\year{2020}\relax
\documentclass[letterpaper]{article} % DO NOT CHANGE THIS
\usepackage{aaai20}  % DO NOT CHANGE THIS
\usepackage{times}  % DO NOT CHANGE THIS
\usepackage{helvet} % DO NOT CHANGE THIS
\usepackage{courier}  % DO NOT CHANGE THIS
\usepackage[hyphens]{url}  % DO NOT CHANGE THIS
\usepackage{graphicx} % DO NOT CHANGE THIS
\usepackage{graphicx}  % DO NOT CHANGE THIS
\usepackage{amsmath,amsfonts,amssymb} % math formula related
\usepackage[linesnumbered,boxed,commentsnumbered, ruled]{algorithm2e} % algorithm related
\usepackage{booktabs,multirow,makecell,threeparttable} % table related
\usepackage{subfig} % figure related
\usepackage{xcolor}

\renewcommand{\vec}[1]{\mathbf{#1}}
\title{Learning Geo-Contextual Embeddings for Commuting Flow Prediction}
\author{Zhicheng Liu\textsuperscript{\rm 1, 2}, Fabio Miranda\textsuperscript{\rm 2}, Weiting Xiong\textsuperscript{\rm 3, 4}, Junyan Yang\textsuperscript{\rm 3}, Qiao Wang\textsuperscript{\rm 1}, Claudio T. Silva\textsuperscript{\rm 2}\\ % All authors must be in the same font size and format. Use \Large and \textbf to achieve this result when breaking a line
\textsuperscript{\rm 1}School of Information Science and Engineering, Southeast University, Nanjing, China\\ 
\textsuperscript{\rm 2}Tandon School of Engineering, New York University, New York, USA\\
\textsuperscript{\rm 3}School of Architecture, Southeast University, Nanjing, China\\
\textsuperscript{\rm 4}Media Lab, Massachusetts Institute of Technology, Boston, USA\\ %If you have multiple authors and multiple affiliations
\{zhichengliu, qiaowang\}@seu.edu.cn, \{fmiranda, csilva\}@nyu.edu, weitingx@mit.edu, yjy-2@163.com % email address must be in roman text type, not monospace or sans serif
}
 
\begin{document}

\maketitle

\begin{abstract}
Predicting commuting flows based on infrastructure and land-use information is critical for urban planning and public policy development.
However, it is a challenging task given the complex patterns of commuting flows.
Conventional models, such as gravity model, are mainly derived from physics principles and limited by their predictive power in real-world scenarios where many factors need to be considered. Meanwhile, most existing machine learning-based methods ignore the spatial correlations and fail to model the influence of nearby regions. To address these issues, we propose \textbf{\underline{G}}eo-contextual \textbf{\underline{M}}ultitask \textbf{\underline{E}}mbedding \textbf{\underline{L}}earner (\textbf{GMEL}), a model that captures the spatial correlations from geographic contextual information for commuting flow prediction. Specifically, we first construct a geo-adjacency network containing the geographic contextual information. Then, an attention mechanism is proposed based on the framework of graph attention network (GAT) to capture the spatial correlations and encode geographic contextual information to embedding space. Two separate GATs are used to model supply and demand characteristics. A multitask learning framework is used to introduce stronger restrictions and enhance the effectiveness of the embedding representation. Finally, a gradient boosting machine is trained based on the learned embeddings to predict commuting flows. We evaluate our model using real-world datasets from New~York~City and the experimental results demonstrate the effectiveness of our proposal against the state of the art.
\end{abstract}

\input{01introduction.tex}

\input{02relatedwork.tex}

\input{03preliminaries.tex}

\input{04methodology.tex}

\input{05experiments.tex}

\input{06conclusion.tex}

\input{07acknowledgement.tex}

\bibliographystyle{aaai}

\bibliography{AAAI-LiuZ.9223}

\end{document}

%% file: 01introduction.tex
\section{Introduction}

% value of commute problem
The commute of people from home to work is a phenomenon that has shaped society and cities throughout the ages, from ancient Egypt to modern New York City~\cite{nyccommute,doi:10.1002/oa.2575}. These daily recurrent movements form a complex network that is highly correlated with the socioeconomic factors of cities~\cite{simini_universal_2012,spadon_reconstructing_2019}. Access to public services, open spaces, transportation and even entertainment all play a role that influence where a worker will live, or where a company will station its offices. 

\input{fig01_overview.tex}

% Importance
In order to have more efficiently planned cities, it is crucial to understand how commuting flows are impacted by infrastructure and land use. This information can be used in urban planning to guide the development of new districts, or in transportation planning to direct the deployment of new modes of transport.
Consider for instance the example of Manhattan, a dense borough of New~York~City. Because of its historical concentration of transit hubs and major corporations, it is the principal destination for workers from the outer boroughs of the city~(see Fig.~\ref{fig:overview - data}).
Redevelopment and rezoning initiatives, however, have contributed to the increase in the number of jobs located in Brooklyn and Queens~\cite{employment}.
Understanding the commuting flow can then help answer many what-if questions in the planning stage, such as \emph{``If a new high-tech industrial park is planned for a region in Brooklyn, from which regions would people commute to work? How should we plan the supporting infrastructure to improve the commuting efficiency?"}, which could help urban planners, policy makers and different stakeholders to make informed decisions. 

% difference from time-series based traffic problem
As such, commuting flow prediction is one of the fundamental problems for urban planning in that it reveals the spatial interactions of supply and demand in a city~\cite{rodrigue_geography_2016}. 
The problem differs from spatio-temporal traffic origin-destination~(OD) forecasting problem. Traffic OD forecasting is essentially a time series prediction problem where the historical movements will be used as input features, while commuting flow prediction problem aims at revealing spatial interaction of supply and demand in a city by predicting the edge-level signals (e.g. the volume of the flow), using \textbf{only} the information of node attributes, such as infrastructure and land use information (see Fig.~\ref{fig:overview - problem}).
Since commuting behaviour shows a daily repeated static pattern, this property enables us to develop a connection between commute and urban indicators.

% limitations: conventional approach and recent machine learning approach

Although the problem has a long history that goes back to the eighteenth century~\cite{monge_memoire_1781}, it remains challenging because of the inherent complexity of cities. Past proposals that use gravity or radiation model~\cite{lenormand_systematic_2016,simini_universal_2012} make simple assumptions of the generation process of commuting flows, and might not capture certain commute patterns. Recently proposed machine learning-based models, such as gradient boosting machine~\cite{pourebrahim_trip_2019}, on the other hand, only consider the features of origin-destination, ignoring the influence of nearby regions.

To address the above issues, we propose a model called Geo-Contextual Multitask Embedding Learner~(GMEL) to predict commuting flows. First, we construct a geographic adjacency network that represents the geographic relationship between regions (Fig.~\ref{fig:overview - solution}). Then, inspired by Tobler's first law of geography~\cite{tobler_computer_1970} which states that \emph{everything is related to everything else, but nearby things are more related than distant things}, we employ Graph Attention Network~(GAT) to encode the geographic contextual information into an embedding space.
Since commuting flow can be viewed as a kind of spatial interaction between origin supply and destination demand~\cite{rodrigue_geography_2016}, we employ two separate GATs to encode the geographic contextual information into two different embedding spaces. This allows us to disentangle the supply and demand characteristics that are hidden in the infrastructure and land use features.
% To disentangle the supply and demand characteristic that are hidden in the infrastructure and land use features, we employ two separate GATs to encode the geographic contextual information into two different embedding space respectively.
To ensure the effectiveness of the embeddings representation, multitask objective functions is used to introduce stronger restrictions forcing the embeddings to encapsulate effective representation for flow prediction~\cite{caruana_multitask_1997}. Finally, we train a gradient boosting machine regression model using the embeddings generated by GMEL as input features.

% contributions
The primary contributions of this paper are the following:

\begin{itemize}
    \item We propose the use of geographic contextual information for commuting flow prediction problem. To our knowledge, we are the first to exploit geographic contextual information in this task.
    
    \item We propose a model (GMEL) to capture the spatial correlations from geographic contextual information and encode the information into embedding space based on graph attention network.
    
    \item We conduct extensive experiments using real-world datasets from New York City. The results demonstrate the effectiveness of our proposed method against the state of the art.
\end{itemize}

%It distinguishes itself from most of the traffic prediction problem. Most of traffic prediction problems can be seen as a time series based prediction problem. However, in this paper, we focus on predicting the edge weights using only the attribute of nodes.

%% file: fig01_overview.tex
\begin{figure}[t!]
    \centering

    \subfloat[]{
        \label{fig:overview - data}
        \begin{minipage}[t]{\linewidth}
            \centering
            \includegraphics[width=\linewidth]{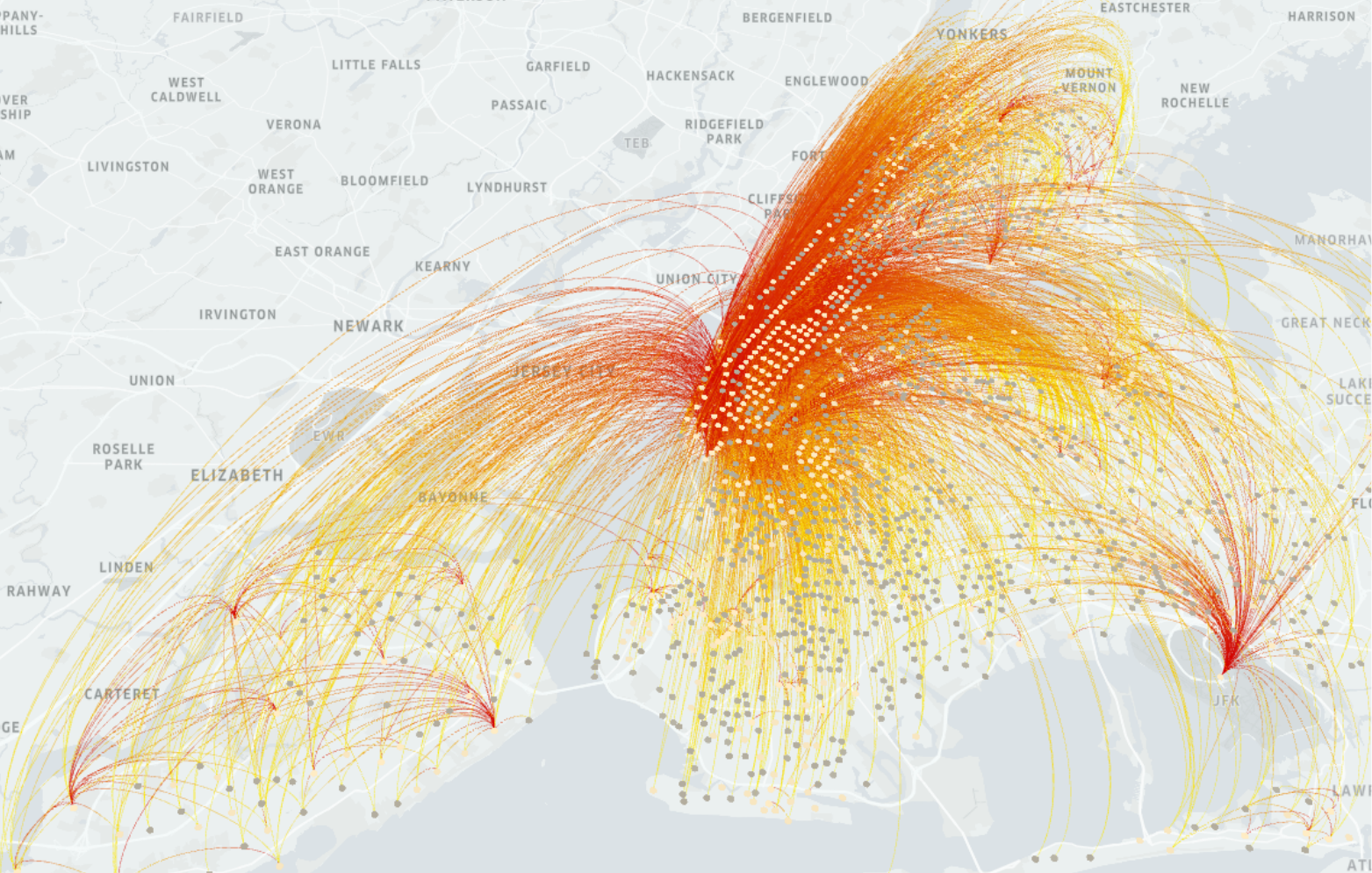}
        \end{minipage}
    }\par
    % \vspace{-0.15in}
    \subfloat[]{
        \label{fig:overview - problem}
        \begin{minipage}[t]{0.5\linewidth}
            \centering
            \includegraphics[width=\linewidth]{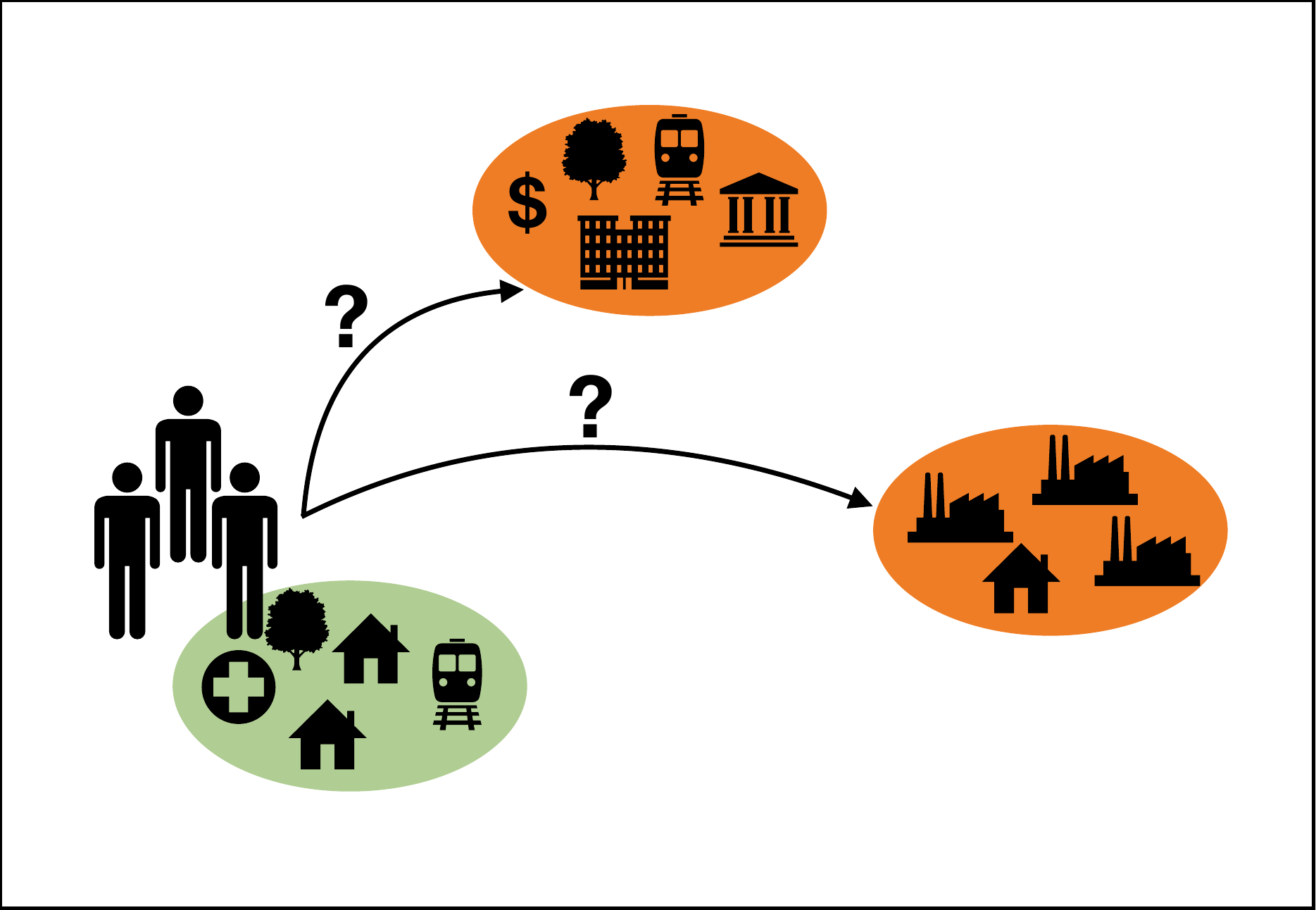}
        \end{minipage}
    }
    \subfloat[]{
        \label{fig:overview - solution}
        \begin{minipage}[t]{0.5\linewidth}
            \centering
            \includegraphics[width=\linewidth]{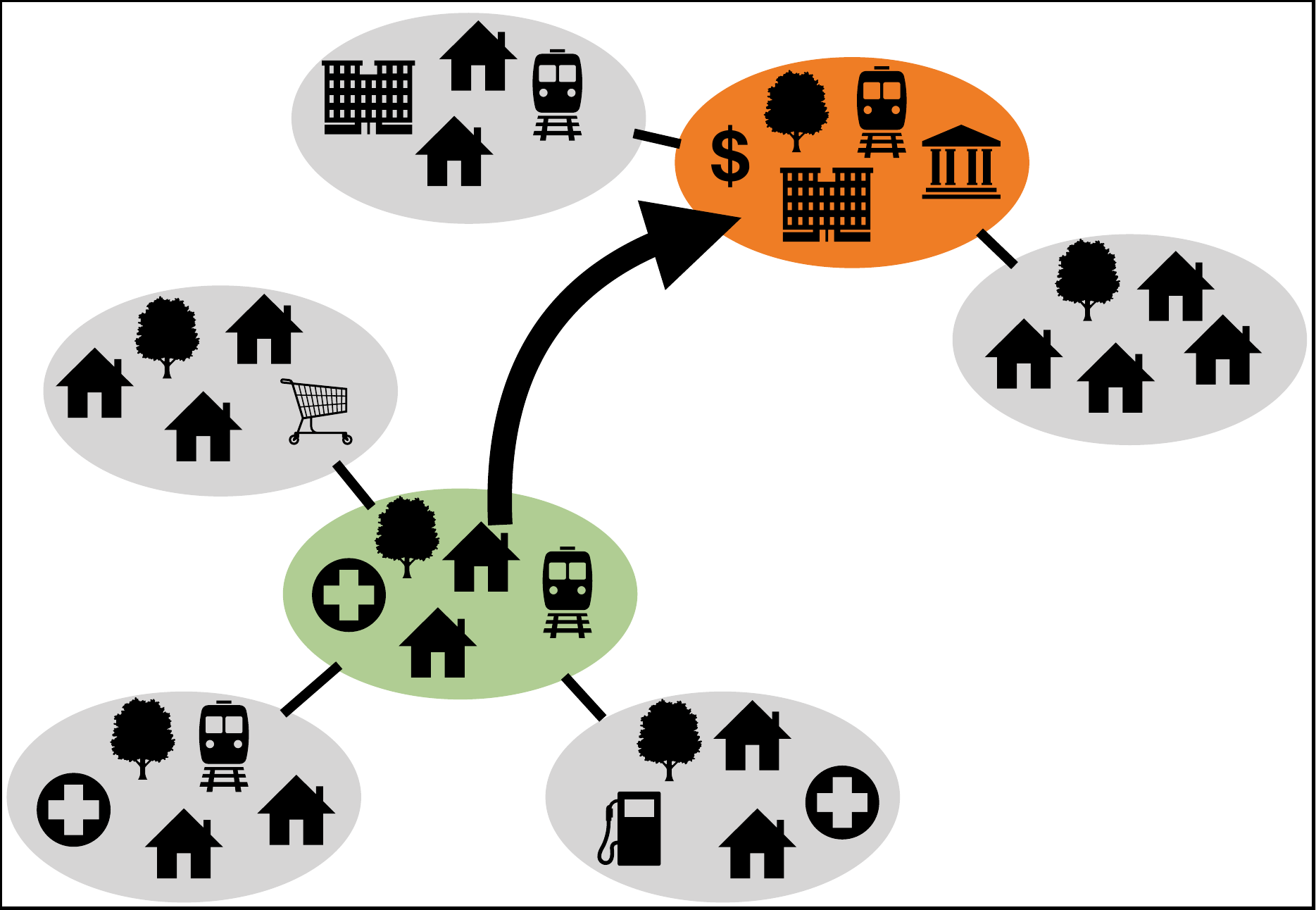}
        \end{minipage}
    }
    % \vspace{-0.1in}
    \caption{Overview. (a) Example of the commuting flow network of New York City in 2015. Yellow indicates origin census geographic units, and red indicates destination units. (b) Illustration of the commuting flow prediction problem when residents choose where to work based on supporting infrastructure, distance, etc. (c) Illustration of our solution that uses geographic contextual information for commuting flow prediction.}
    \label{fig:overview}
\end{figure}

%reflecting the spatial interaction of supplies and demands in city.

%% file: 02relatedwork.tex
\section{Related Work}

\subsection{Commuting Flow Prediction}
% essence of the problem and difference with traffic OD forecasting
In this paper, we focus on commuting flow prediction problem~\cite{spadon_reconstructing_2019}. It is important to note that the problem formulation is different from the traffic origin-destination~(OD) forecasting problem~\cite{xiong_dynamic_nodate,iwata}. Although both of the problems view human movements as networks, traffic OD forecasting problem is essentially an edge-level time series prediction problem where the historical network edges (e.g. ODs) can be the input features of the model~\cite{wang_origin-destination_2019}, while commuting flow prediction problem aims at predicting the edge weights (e.g. the volume of the flow) utilizing only the attribute of nodes.

% gravity model, radiation model and intervening opportunity model
Gravity model~\cite{lenormand_systematic_2016} is a widely used conventional model for commuting flow prediction which makes simple assumptions of the generation process of commuting flow. For example, the original form of gravity model~\cite{zipf_p1_1946} assumes the number of commuters traveling from one region to another is proportional to the product of the population of origin and destination and decays with the distance of the trip, as shown below:

\vspace{-0.2in}
\begin{align}
    \hat{T}_{ij} = \alpha_0 \dfrac{M_i^{\alpha_1} M_j^{\alpha_2}}{d_{ij}^{\alpha_3}}
\end{align}

\noindent where $M_i$ and $M_j$ are the population of region i and j respectively, $d_{ij}$ is the travel distance between the two regions, and $\{\alpha_0, \alpha_1, \alpha_2, \alpha_3\}$ are the parameters to be estimated. However, this simple assumption might not capture the complex nature of a city~\cite{albeverio_dynamics_2007}.

Recently, researchers have used nonparametric models, such as gradient boosting machine, to capture the complex nature of spatial interaction represented by commuting flow~\cite{pourebrahim_trip_2019,robinson_machine_2018}. These off-the-shelf machine learning models usually present better performance than conventional physics-derived models. However, these models simply use origin-destination node attributes as input features to fit a regression model, ignoring the influence of nearby regions. Another family of conventional models is called intervening opportunity model. These models consider the influence of nearby potential competitors of origin or destination, such as radiation model~\cite{simini_universal_2012,yang2014limits}. Inspired by the idea of intervening opportunity, we propose using the geographic contextual information to develop the regression model where the embeddings of each node is encoded with the influence of nearby regions.
%

%These models are validated to be effective in the nation-wide scale and county level spatial resolution. However, in a more practical city-wide scale and finer spatial resolution, these simple assumptions might not capture the complex nature of city\cite{}. Another physics principle based model\cite{}

% machine learning models

\subsection{Graph Representation Learning}
% GNN: GraphSAGE and GAT
Several graph representation learning methods have been proposed recently. A general inductive framework called GraphSAGE is proposed by~\cite{hamilton_inductive_2017}, which leverages node attribute to generate node embeddings in a message-passing way. Also, graph attention network~\cite{velickovic_graph_2017} leverages self-attention mechanism to allow messages passed by neighbors to be aggregated with different weights. Motivated by these works, we use the framework of graph attention network and adapt the attention mechanism to our tasks so that our model could capture the geographic context.

% GNN for Commuting Flow
Several applications have also been proposed based on graph neural network. \cite{pan_urban_2019} utilized a GAT-like structure to learn embeddings that capture spatial correlations of traffic patterns. \cite{wang_origin-destination_2019} proposed a GraphSAGE-like graph embedding model to capture the spatial mobility patterns and neighboring correlations. \cite{zhang_flow_2019} proposed a multitask learning framework to simultaneously predict the node and edge traffic flows.
%
% \highlight{\cite{iwata} aims at predicting the dynamic nearby regional transitions of the population in the city based on the previous time slot population. Our goal is different, as we aim to predict the city-wide temporally static transitions of the population based on geographical features.}
%
Few studies explored the use of graph neural network to capture spatial correlations for commuting flow predictions.
%
% \highlight{\cite{Wang:2017:RRL:3132847.3133006} proposes a spatial auto-regression, which doesn't fit our scenario since we aim at predicting commuting flows using only the infrastructure and land use information.}

% Our work differentiates with the above work in that we focus on modeling commuting flows which is daily recurrent movements that itself reflect the 
% based on infrastructure and land use information. (2) We further propose an agglomerative process to make an inference of housing market structure.

% In this paper, however, we aim at propose a geo-contextual embedding learner to capture the spatial correlations . \fabio{What sets us apart? None of them consider geo-contextual information? If so, it should written here.}

%% file: 03preliminaries.tex
\section{Preliminaries}

In this section, we introduce the definitions and problem formulation.

\textit{Definition 1} \textbf{Urban Geographic Unit}: We partition the city into $N$ urban geographic units $v_1, v_2, ..., v_N$. The geographic units can be street blocks, census tracts, zip code areas, etc. 

\textit{Definition 2} \textbf{Urban Indicators}: The urban indicator $\vec{a}_i$ is a vector that serves as the attribute of $v_i$. It characterizes the aggregated information of infrastructure and land use of the geographic units.

\textit{Definition 3} \textbf{Geo-Adjacency Network}: The Geo-adjacency network is an undirected weighted graph $G_{adj}=(V, E, A)$ where $V = \{v_1, v_2, ..., v_{N}\}$ is the set of urban geographic units which serves as the nodes of the graph, $E = \{e_{ij} | \text{$v_i$, $v_j$ are geographically adjacent}, 1\leq i, j\leq N\}$ is the set of edge features that describes the strength of correlations (e.g. travel distance, trip duration) and $A = \{\vec{a}_1, \vec{a}_2, ..., \vec{a}_N\}$ is the set of urban indicators that serves as the node attributes.

In our case, we use census tracts as the urban geographic unit. The geo-adjacency network of New York City is shown in Fig. \ref{fig:geo_adj_net}.

\input{geo_adj_network.tex}

\textit{Definition 4} \textbf{Distance Matrix}: The distance matrix $D$ is a $N$-by-$N$ matrix where the entry $D_{ij}$ represents the travel distance from $v_i$ to $v_j$.   

\textit{Definition 5} \textbf{Commuting Trips}: Commuting trips are a set of triplets $T = \{(v_i, v_j, T_{ij})\}$ where $v_i$ is the trip origin node, $v_j$ is the trip destination node, $T_{ij}$ is the commuting flow, i.e. the number of commuters travel from $v_i$ to $v_j$. Note that $T_{ij}$ can be seen as an edge-level flow. We also define two kinds of node-level flows, i.e. in-flow and out-flow. We denote the out-flow as $T_{i:}$ representing the total number of outgoing commuters from $v_i$ and denote the in-flow as $T_{:j}$ representing the total number of incoming commuter to $v_j$.

\textbf{Problem}: Given $G_{adj} = (V, E, A)$ and $D$, develop a regression model to predict $T_{ij} \in T$.

%% file: geo_adj_network.tex
\begin{figure}[hbtp]
    \centering
    \resizebox{.8\linewidth}{!}{
    \includegraphics[width=0.7\linewidth]{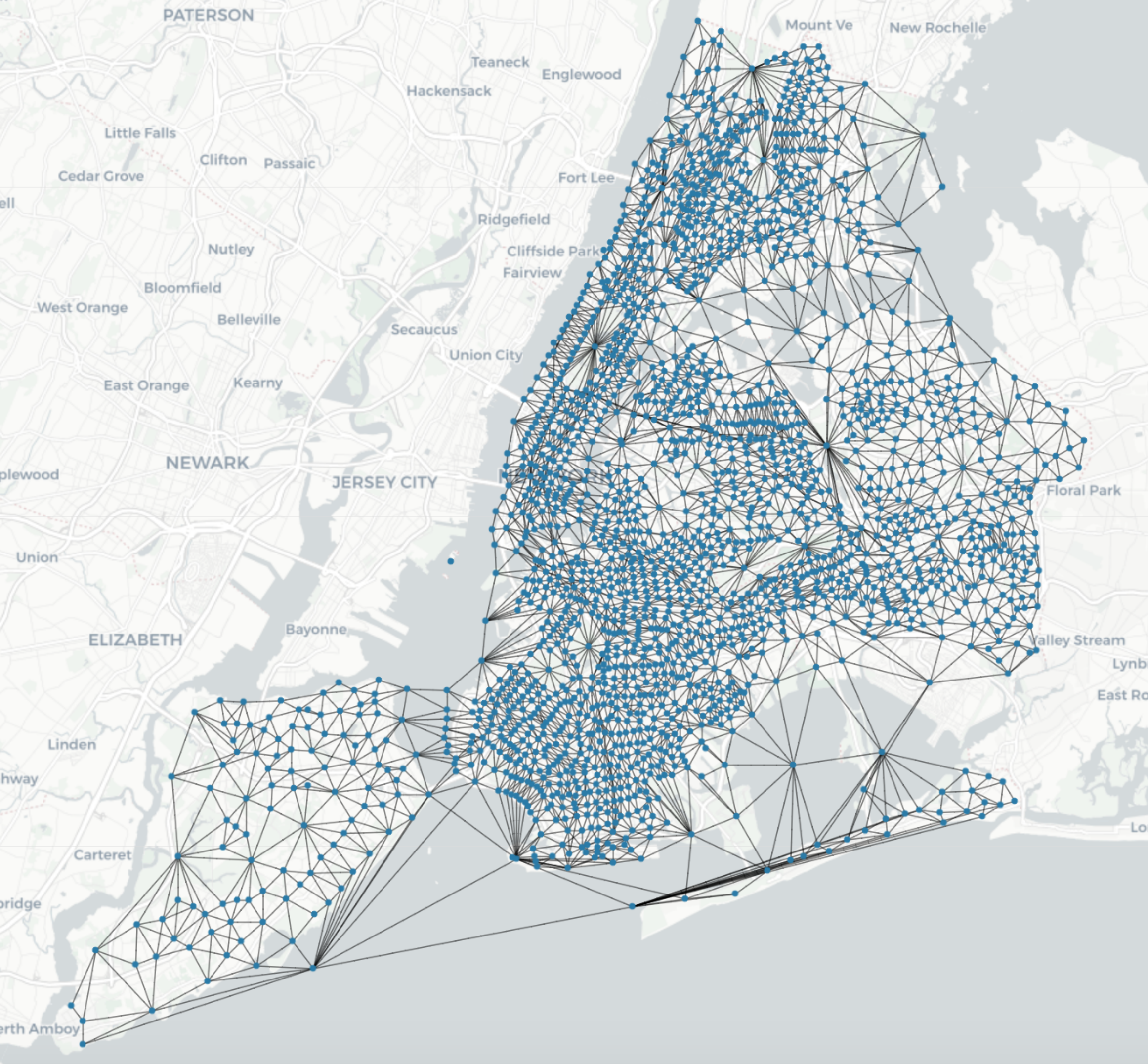}
    }
    \caption{Geo-adjacency network of New York City. The dots represent the centroids of census tracts and the lines represent the edges.}
    \label{fig:geo_adj_net}
\end{figure}

%% file: 04methodology.tex
\section{Methodologies}

In this section, we describe the architecture of our model for commuting flow prediction. Basically, our model consists of two components: Geo-contextual Multitask Embedding Learner and Flow Predictor.

1) \textbf{\textbf{\underline{G}}eo-contextual \textbf{\underline{M}}ultitask \textbf{\underline{E}}mbedding \textbf{\underline{L}}earner (\textbf{GMEL})}. GMEL is designed to capture the spatial correlations from geographic context. Basically, the geographic context can be viewed as the graph neighborhoods of $G_{adj}$. GMEL utilizes Graph Attention Network~(GAT) to encode the geographic contextual information into an embedding space. To disentangle the supply and demand characteristics that are hidden in infrastructure and land use, GMEL employs two separate GATs to encode the geographic contextual information into two different embedding space. To ensure the effectiveness of the embeddings representation, GMEL employs multitask learning framework which imposes stronger restrictions forcing the embeddings to encapsulate effective representation for flow prediction~\cite{caruana_multitask_1997}.

2) \textbf{Flow Predictor}. Considering the learned embeddings from GMEL, we employ gradient boosting machine~(GBM) as the regression model to predict commuting flows. GBM has advantages in handling dense numerical features~\cite{ke_deepgbm:_2019}, such as the learned embeddings in our scenarios. By iteratively evaluating the largest information gain of features, GBM can automatically select and combine useful numerical features to fit the targets~\cite{friedman_elements_2001}. This is why most recently proposed machine learning models for commuting flow prediction employ gradient boosting regression tree~(GBRT) or random forest as the regression function~\cite{spadon_reconstructing_2019,pourebrahim_trip_2019,robinson_machine_2018}. In particular, we use GBRT in this paper.

\subsection{Framework}

\input{embedding_framework.tex}

The framework of GMEL is shown in Fig.~\ref{fig:emb learning framework}. GMEL aims at learning effective embeddings of urban geographic units which encode the geographic contextual information. To learn the supply  and demand characteristics for each geographic unit respectively, we employ two separate GATs to encode this information. The generated embeddings are then applied to a bilinear function to predict the flow. Meanwhile, these embeddings will also be applied to two linear functions to predict the in/out-flow of the geographic units. The overall prediction loss is the weighted sum of the three tasks' loss, and we use backpropagation to train GMEL in an end-to-end manner.

\subsection{Graph Attention Network}

Graph attention network~(GAT) iteratively aggregates the information from node neighborhood and updates the node states with nonlinearity. The weight used to aggregate the neighborhood messages depends on the features of two connecting nodes and edge features.

Assume the state of node $i$ is $h_i^{(l)} \in \mathbb{R}^{m \times 1}$ in the $l$-th layer and the features of edge $(v_i, v_j)$ is $e_{ij} \in \mathbb{R}^{n \times 1}$. GAT first applies linear transformation to these vectors.

\vspace{-0.1in}
\begin{align}
    &z_i^{(l)} = W^{(l)} h_i^{(l)} \\
    &c_{ij}^{(l)} = V^{(l)} e_{ij}
\end{align}

\noindent where $W^{(l)} \in \mathbb{R}^{k \times m}$ and $V^{(l)} \in \mathbb{R}^{t \times n}$ are trainable parameter matrices. The resulting $z_i$ is the message vector passed to neighbors. Before aggregating these message vectors, an attention score for each edge is calculated:

\vspace{-0.1in}
\begin{align}
    &r_{ij}^{(l)} = \sigma(a^{(l)T}(z_i^{(l)}||c_{ij}^{(l)}||z_j^{(l)})) 
\end{align}

\noindent where $\sigma(\cdot)$ is a nonlinear function (e.g. Relu, Sigmoid), $a^{(l)} \in \mathbb{R}^{(2k+t) \times 1}$ is a trainable parameter vector that maps the concatenation of messages into a scalar value and $||$ denotes the concatenation operation. Then, the attention scores are normalized by softmax:

\vspace{-0.1in}
\begin{align}
    &\alpha_{ij}^{(l)} = \dfrac{exp(r_{ij}^{(l)})}{\sum_{k\in\mathcal{N}(i)}exp(r_{ik}^{(l)})}
\end{align}

\noindent where $\mathcal{N}(i)$ denotes the graph neighborhood of the $i$-th node. The final aggregation process consists of two parts representing the neighborhood impact and self impact respectively:

\vspace{-0.1in}
\begin{align}
    &h_i^{(l+1)}=\sigma(\sum_{j\in\mathcal{N}(i)}\alpha_{ij}^{(l)}z_j^{(l)} + U^{(l)}h_i^{(l)})
\end{align}

\noindent where $U^{(l)} \in \mathbb{R}^{k \times m}$ is a trainable parameter matrix and $\mathcal{N}(i)$ is the neighborhood of node $i$. 

\subsection{Modeling Supply and Demand Characteristic}

Commuting flows can be viewed as a kind of spatial interactions between supplies and demands~\cite{rodrigue_geography_2016}. Our model holds an underlying assumption that the flows are determined by the supply characteristic of the origin geographic unit and the demand characteristic of the destination geographic unit.

To model both supply and demand characteristic of each geographic unit, we use two separate GATs. In Fig.~\ref{fig:emb learning framework}, the $\text{GAT}^{\text{(org)}}$ extracts demand characteristic from origin geographic units and encodes the characteristic into origin embeddings. The $\text{GAT}^{\text{(dst)}}$ extracts supply characteristic from destination geographic units and encodes the characteristic into destination embeddings. The two GATs have the same structure, but the attention mechanism in GAT will assign different weights to different features based on the origin or destination roles, thus modeling supply and demand characteristics.

%That's the reason why we use GAT rather than Graph Convolution Network(GCN)\cite{schlichtkrull_modeling_2018} which simply takes the average of the neighbors' messages. 

\subsection{Multitask Learning}

As the goal of GMEL is to learn embeddings that encode supply and demand characteristic for commuting flow prediction, we adopt multitask learning framework to put more restrictions for the GMEL training process.

\subsubsection{Main Task: Predicting Commuting Flow}

Having the origin and destination embeddings $h_i^{(org)}$ and $h_j^{(dst)}$ from GATs, a bilinear model is used to predict the commuting flow:

\vspace{-0.1in}
\begin{align}
    \hat{T}_{ij} = h_i^{(org)T} W_{b} h_j^{(dst)}
\end{align}

\noindent where $W_{b} \in \mathbb{R}^{m \times m}$ is a trainable parameter matrix modeling the interactions between origin embeddings and destination embeddings. The corresponding loss function of the main task is:

\vspace{-0.1in}
\begin{align}
    \mathcal{L}_{main} = \dfrac{1}{|T|} \sum_{i,j} (\hat{T}_{ij} - T_{ij})^2
\end{align}

\noindent where $|T|$ is the total number of trips.

\subsubsection{Subtasks: Predicting In/Out Flow}

We include prediction of the in/out-flow as two subtasks, i.e. predicting the total number of incoming/outgoing commuters of each geographic unit. The intuition is that the commuting flows and in/out-flows are highly correlated and, thus, the two subtasks would impose stronger restrictions on the training process of GMEL. The in/out-flows are predicted by linear functions:

\vspace{-0.1in}
\begin{align}
    \hat{T}_{i:} = \vec{w}_{out}^T h_i^{(org)} \\
    \hat{T}_{:j} = \vec{w}_{in}^T h_j^{(dst)}
\end{align}

\noindent where $\vec{w}_{out}, \vec{w}_{in} \in \mathbb{R}^{m \times 1}$ are trainable vector parameters. The corresponding loss function of two subtasks are:

\vspace{-0.1in}
\begin{align}
    \mathcal{L}_{out} = \dfrac{1}{N} (\hat{T}_{i:} - T_{i:})^2 \\
    \mathcal{L}_{in} = \dfrac{1}{N} (\hat{T}_{:j} - \hat{T}_{:j})^2
\end{align}

\noindent where $N$ is the number of geographic units.

\subsubsection{Overall Loss Function}

The overall loss function of GMEL is formulated as the weighted sum of all three tasks:

\vspace{-0.1in}
\begin{align}
    \mathcal{L}_{GMEL} = \lambda_{main} \mathcal{L}_{main} + \dfrac{\lambda_{sub}}{2} (\mathcal{L}_{in} + \mathcal{L}_{out})
    \label{eq:overall loss}
\end{align}

\noindent where $\lambda_{main}$, $\lambda_{sub}$ are the hyperparameters representing the weights for main task and subtasks respectively.

\subsection{Training Algorithm}

Recall that our model consists of two components: GMEL and flow predictor. We train GMEL using stochastic gradient descent method in an end-to-end manner. The learning process of GMEL can be seen as pre-training. Having the embeddings from well-trained GMEL, a GBRT is trained as flow predictor based on the concatenation of origin-destination embeddings and travel distance to predict the commuting flow. The training process is summarized in Algorithm \ref{alg:learning process}.

\begin{algorithm}[t]
    \caption{Training Algorithm}
    \SetKwData{Index}{Index}
    \label{alg:learning process}
    \KwIn{Geo-adjacency Network $G_{adj} = (V, E, A)$, \\
    Distance Matrix $D$, \\
    Commuting Trips $T_{train}=\{(v_i, v_j, T_{ij})\}$}
    \BlankLine
    \KwOut{The learned GMEL,\\
    The learned flow predictor $\hat{f}$}
    \BlankLine

    \tcc{GMEL Learning}
    
    \Repeat{stopping criterion is met}{
        $T_{batch} \leftarrow$ Draw a training batch from $T_{train}$
        
        $\{h_i^{(org)}\} \leftarrow GAT^{(org)}(G_{adj})$
        
        $\{h_j^{(dst)}\} \leftarrow GAT^{(dst)}(G_{adj})$
        
        Evaluate $\mathcal{L}_{GMEL}$ by $(\{h_i^{(org)}\}, \{h_j^{(dst)}\}, T_{batch})$ using Equation \ref{eq:overall loss}
        
        $\nabla \mathcal{L}_{GMEL} \leftarrow \text{Backpropagate } \mathcal{L}_{GMEL}$
        
        $w \leftarrow w - \gamma \nabla \mathcal{L}_{GMEL}$ \tcp{$\gamma$ is the learning rate}
    }
    
    \BlankLine
    \tcc{Flow Predictor Learning}
    
    $\{h_i^{(org)}\} \leftarrow GAT^{(org)}(G_{adj})$
    
    $\{h_j^{(dst)}\} \leftarrow GAT^{(dst)}(G_{adj})$
    
    $\mathcal{X}_{input} \leftarrow \{\}$, 
    $\mathcal{Y}_{input} \leftarrow \{\}$
    
    \For{$(v_i, v_j, T_{ij}) \text{ in } T_{train}$}{
        $\mathcal{X}_{input} \leftarrow \mathcal{X}_{input} \cup Concat(h_i^{(org)}, h_j^{(dst)}, D_{ij})$
        
        $\mathcal{Y}_{input} \leftarrow \mathcal{Y}_{input} \cup T_{ij}$
    }
    
    \BlankLine
    $\hat{f} \leftarrow \text{Train GBRT on } (\mathcal{X}_{input}, \mathcal{Y}_{input})$

\end{algorithm}

%% file: embedding_framework.tex
\begin{figure}[!t]
    \centering
    \includegraphics[width=1.0\linewidth]{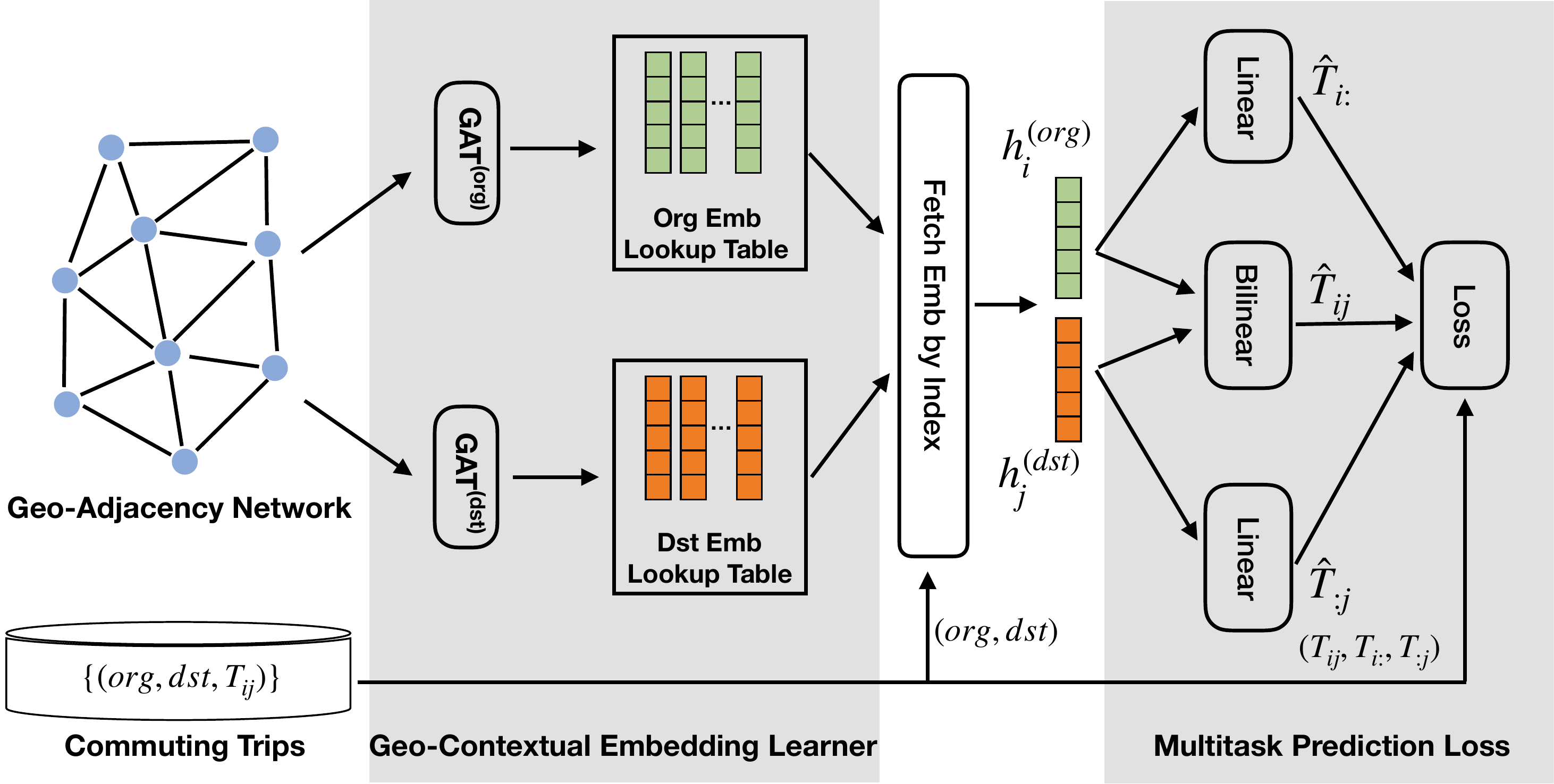}
    \caption{Framework of GMEL}
    \label{fig:emb learning framework}
\end{figure}

%% file: 05experiments.tex
\section{Experiments}

In this section, we provide an empirical evaluation of our proposed model on real-world dataset.

\subsection{Datasets}

We validate our proposed model on real-world datasets from New~York~City. To make comparison with state-of-the-art models in the literature, we use similar experimental settings as reported in \cite{pourebrahim_trip_2019}. We use the 2010 New~York~City census tracts as geographic units~(2168 units in total). For commuting trips and urban indicators, we use the following datasets:

\subsubsection{LODES}

The 2015 Origin-Destination Employment Statistics~(LODES) dataset presents the commuting trips of interest~\cite{lehd}. It is collected yearly and records the home and employment locations of workers, representing stable commuting flows.
These flows are aggregated into geographic unit level flow. 3,031,641~commuters and 905,837~pairs of origin-destination trips were collected in New~York~City. We randomly divide the commuting trips into training, validation and test datasets by 6:2:2.

\subsubsection{PLUTO}
The 2015 NYC Primary Land Use Tax Lot Output~(PLUTO) presents the urban indicators of interest~\cite{pluto}. It records land use and infrastructure information at the tax lot level. This information is aggregated into census tract level (65 urban indicators for each census tract). A summary of the urban indicators is listed in Table~\ref{tab:urban_metrics}.

\subsubsection{OSRM}
We employ Open Source Routing Machine~(OSRM) to measure the travel distance between the centroids of census tracts~\cite{luxen_real-time_2011}. The travel distances will serve as the edge features of the geo-adjacency network.

\input{urban_metrics.tex}

\subsection{Baselines}

To show the effectiveness of our model, we compare our model with the following baselines:

\begin{itemize}
    \item \textbf{Gravity Model with Power-Law Decay~(GM-P)}: Gravity model with power-law distance decay function is the most classic model for spatial interaction model. It's widely used in predicting commuting flows, cargo shipping volume, etc. Basically, gravity model is a log-linear model. The difference of models in this family lies in the form of the distance decay function. For further details of gravity model, we refer the readers to~\cite{lenormand_systematic_2016}.
    
    \item \textbf{Gravity Model with Exponential Decay~(GM-E)}: Gravity model with exponential distance decay function is another model in gravity model family. It is reported to have better performance in predicting commuting flows~\cite{lenormand_systematic_2016}.
    
    \item \textbf{Random Forest~(RF)}: Recently, researchers proposed to use gradient boosting machine to predict commuting flows. RF is reported as the state-of-the-art model~\cite{pourebrahim_trip_2019,spadon_reconstructing_2019}.
    
    \item \textbf{Gradient Boosting Regression Tree~(GBRT)}: GBRT belongs to gradient boosting machine family and is widely used as regression model.
    
    \item \textbf{Node2vec}: Node2vec is an unsupervised learning model to learn node embeddings from graph structured data~\cite{grover_node2vec:_2016}. Recently, its variant have also been applied to learn location embeddings for spatio-temporal prediction tasks~\cite{Wang:2017:RRL:3132847.3133006}. We incorporate Node2vec to learn the embeddings for each census tract on the geo-adjacency network and use these embeddings as inputs to train a gradient boosting regression tree as flow predictor.

\end{itemize}

To validate the effectiveness of our model architecture, we also implement two variants of our model:

\begin{itemize}
    \item \textbf{GMEL-noMul}: We remove multitask settings and only keep the main task, i.e. setting $\lambda_{main} = 1, \lambda_{sub} = 0$ in Equation \ref{eq:overall loss}.
    
    \item \textbf{GMEL-noSep}: We remove the settings of using two separate GATs to model supply and demand characteristics respectively. Instead, only one GAT is used to generate embeddings and this set is used for both origin and destination embeddings. 
\end{itemize}

The above baseline models are compared to the GMEL with multitask weights to be $\lambda_{main} = 0.5, \lambda_{sub} = 0.5$, the embeddings size of both GATs to be 128, and number of GAT layers to be 2.

We implemented our model and the baselines using PyTorch~\cite{paszke2017automatic} and Deep Graph Library~\cite{wang2019dgl}. The experiments were executed on a Intel E5-2690~v4 2.6~GHz, 256~GB of RAM, and a NVIDIA~Tesla~P100~GPU with 12~GB of RAM.

\subsection{Evaluation Metrics}

To measure the prediction performance, we adopt three evaluation metrics: Root Mean Square Error~(RMSE), Mean Absolute Error~(MAE) and Common Part of Commuters~(CPC).

\begin{align}
    RMSE &= \sqrt{ \dfrac{1}{|T|} \sum_{i,j} (\hat{T}_{ij} - T_{ij})^2 } \\
    MAE &= \dfrac{1}{|T|} \sum_{i,j} |\hat{T}_{ij} - T_{ij}| \\
    CPC &= \dfrac{2\sum_{ij}min(\hat{T}_{ij}, T_{ij})}{\sum_{ij}\hat{T}_{ij} + \sum_{ij}T_{ij}}
\end{align}

RMSE and MAE are widely used as evaluation metrics for regression problem. CPC is widely used in commuting flow prediction problem~\cite{lenormand_systematic_2016,robinson_machine_2018}, and it measures the common part of agreements between predicted value and target value. CPC is 0, when no agreement is found, and is 1, when the two are identical.

\input{hyperparameters_sensitivity.tex}

\subsection{Performance Analysis}

\input{performance.tex}

We evaluate the performance of the baseline models and our model on the test set, and summarize the results in Table~\ref{tab:performance}. From the experiments, we have the following observations:

\begin{itemize}
    \item Gravity models have the worst performance among all models. The reason might be that simple assumptions of gravity model cannot capture the complex patterns of commuting flows and thus lead to poor predictive power.
    
    \item RF and GBRT generally have better performance than gravity models, which is in accordance with recently published literature~\cite{pourebrahim_trip_2019}. The reason might be that gradient boosting machine is better capable of handling nonlinearity. %Node2vec performs better than RF, but worse than GBRT.
    
    \item Node2vec is slightly better than RF and only comparable to GBRT, even though it uses graph neighborhood structure to generate embeddings. The reason might be Node2vec is designed to preserve network neighborhood of nodes, but this neighborhood information is not useful for characterizing the supply and demand characteristics of the city.
    
    \item All GMEL variants outperform the above baseline models. This verifies the effectiveness of leveraging geographic contextual information for commuting flow prediction.
    
    \item GMEL outperforms GMEL-noMul and GMEL-noSep. This shows the effectiveness of multitask learning framework and the necessity of modeling supply and demand characteristic separately.
\end{itemize}

To this end, we have validated the effectiveness of our model.

\subsection{Residual Analysis}

\input{spatial_errors.tex}

To illustrate the effectiveness of exploiting spatial correlations, we present the residual maps in Fig. \ref{fig:spatial error}. These maps show the difference between predicted and ground-truth incoming flows, i.e. the sum of the residuals of flows to the same destination, in each census tract. We compare GMEL with the state-of-the-art model GBRT~ \cite{pourebrahim_trip_2019}. In Fig. \ref{fig:spatial error}, we can observe that the residuals of GMEL are spatially smoother than that of GBRT. The reason is that GMEL exploits geographic contextual information to capture spatial correlations, and in doing so the prediction will take into account both the characteristics of regions of interest and the influence of nearby regions.

\input{feature_saliency.tex}

\subsection{Parameter Sensitivity Analysis}

We also analyze the parameter sensitivity of our model. Three main hyperparameters of GMEL are examined, namely the number of GAT layers, embedding size and multitask weights. The results are shown in Fig. \ref{fig:param sensitivity}.

\subsubsection{The effect of number of GAT layers}

The number of GAT layers determines the depth of graph neighborhood. For example, if the number of GAT layers is 2, then all graph neighboring nodes within two-hops would have an effect on the target node. In our scenario, this hyperparameter implicitly defines the geographic range of influence. From Fig.~\ref{fig:hyper-layers}, we can see that when the number of GAT layers is one, the model performs worse. When the number of GAT layers is greater than or equal to two, the performance doesn't fluctuate too much. This indicates that the effective graph neighborhood is two-hops graph neighborhood. In New York City, the two-hops graph neighborhood covers on average 1.5~km, which is approximately 15-minutes walking distance.

\subsubsection{The effect of embedding size}

We also conduct experiments on several alternatives of embedding size, i.e. 32, 64, 128, 256. In Fig. \ref{fig:hyper-embsize}, we can find that the performance increases as the embedding size increases from 32 and saturates at the size of 128. 

\subsubsection{The effect of multitask weights}

Different set of multitask weights are also tested, see Fig. \ref{fig:hyper-multitask}. Recall that the subtasks are introduced to enhance the performance of the main task. Indeed, we can observe in Fig. \ref{fig:hyper-multitask} that when the weight of subtasks increases, the performance of the main task keeps increasing until the weights of the main task and subtasks are equal, i.e. $\lambda_{main} = 0.5, \lambda_{sub} = 0.5$.

\subsection{Feature Sensitivity Analysis}

We evaluate the impact of the urban indicators by computing the saliency map of GMEL~\cite{simonyan2013deep}. In our case, saliency map represents the average gradients of output estimates with regards to urban indicators, exhibiting its overall effect on commuting flows. A larger absolute value of the saliency map points to a more prominent urban indicator. Three saliency maps are evaluated: edge-level flows, in-flows and out-flows.
Fig.~\ref{fig:saliency} shows the most prominent urban indicators.
These salient urban indicators present the supply and demand characteristics for different kind of flows. For example, the number of buildings per square meters, indicating job opportunities, is salient for in-flow, meanwhile, floor area ratio of residence, indicating the density of regular residences, is salient for out-flow.
%
%For edge-level flows, the top three salient urban indicators are average built floor area ratio, the number of loft buildings and the number of buildings per square meters.
%
%For in-flows, the top three salient urban indicators are the number of miscellaneous buildings (e.g. court house, public parking area), the number of condominiums and the number of tax lots of open space and outdoor recreation land use type.
%
%For out-flows, the top three salient urban indicators are the number of tax lots of open space and outdoor recreation land use type, the number of loft buildings and the number selected government installations buildings (e.g. Fire Department, Police Department, Department of Real Estate).

\subsection{Case Study}

To further evaluate the usefulness of our proposal, we show a case study focusing on census tracts that experienced major changes in their urban indicators between the years of 2013 and 2015.
We first select a set of 5 census tracts that had the largest changes when considering their urban indicators between these two years.
Next, we train a model on the 2013 data set (PLUTO and LODES) and test the prediction performance of the model considering the 2015 data set. In our experiments, the mean absolute error and standard deviation between the predicted flow values and the groundtruth for the selected 5 census tracts are the following: $1.07 \pm 1.25, 1.40 \pm 2.12, 1.26 \pm 2.08, 0.99 \pm 1.73, 0.86 \pm 1.37$.
By having a model trained on a particular year, GMEL can be used to predict the origin and destination of new commuting flows, given changes in the urban indicators.
This highlights how our proposal can guide urban planners and policy makers to make informed decisions when it comes to new urban development scenarios.

%% file: urban_metrics.tex
% Table generated by Excel2LaTeX from sheet 'features'
\begin{table}[t!]
  \centering
  \caption{Summary of Urban Indicators}
    \begin{tabular}{ccp{.45\linewidth}}
    \toprule
    \textbf{Categories} & \textbf{\# Features} & \multicolumn{1}{c}{\textbf{Contents}} \\
    \midrule
    \midrule
    \multirow{6}[0]{*}{Infrastructure} & \multirow{6}[0]{*}{40}    & The number of different types of buildings~(25), the density of commercial/residential/etc. units~(4), the number of buildings in each built year interval~(11) \\
    \midrule
    \multirow{6}[0]{*}{Land Use} & \multirow{6}[0]{*}{23}    & The number of tax lots in different land use~(11), the land area ratio of retail/office/etc.~(10), statistics of floor area ratio~(2) \\
    \midrule
    \multirow{3}[0]{*}{Speciality} & \multirow{3}[0]{*}{2}     & Whether or not the census tract contains landmarks or historic districts~(2) \\
    \midrule
    \multirow{1}[0]{*}{Total} & \multirow{1}[0]{*}{65}    &  \\
    \bottomrule
    \end{tabular}%
  \label{tab:urban_metrics}%
\end{table}%

%% file: hyperparameters_sensitivity.tex
\begin{figure*}[ht!]
    \centering
    \subfloat[Effect of number of GAT layers.]{
        \begin{minipage}[t]{0.33\linewidth}
            \centering
            \includegraphics[width=1.0\linewidth]{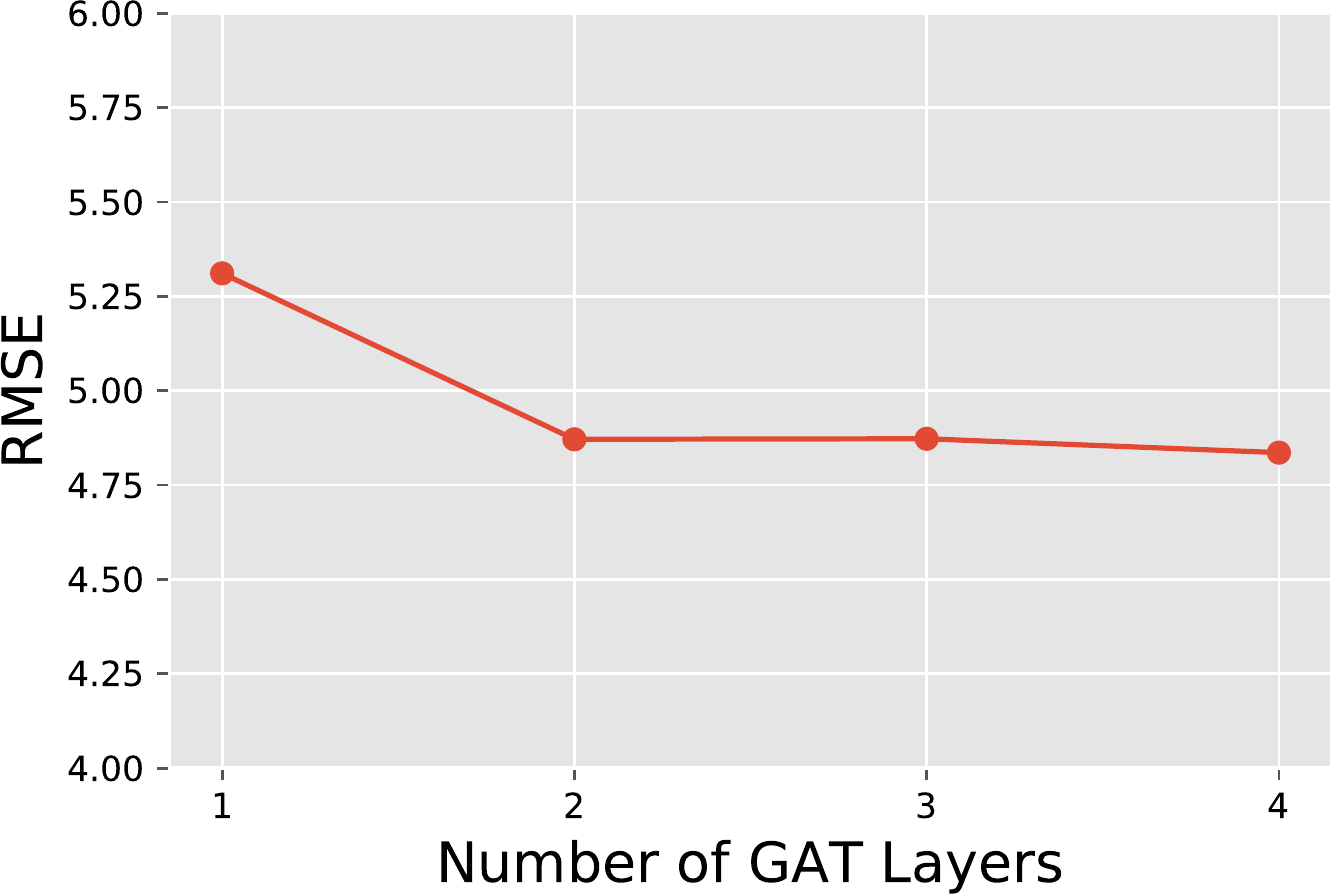}\\[-0.25in]
            \label{fig:hyper-layers}
        \end{minipage}
    }
    \subfloat[Effect of embedding size.]{
        \begin{minipage}[t]{0.33\linewidth}
            \centering
            \includegraphics[width=1.0\linewidth]{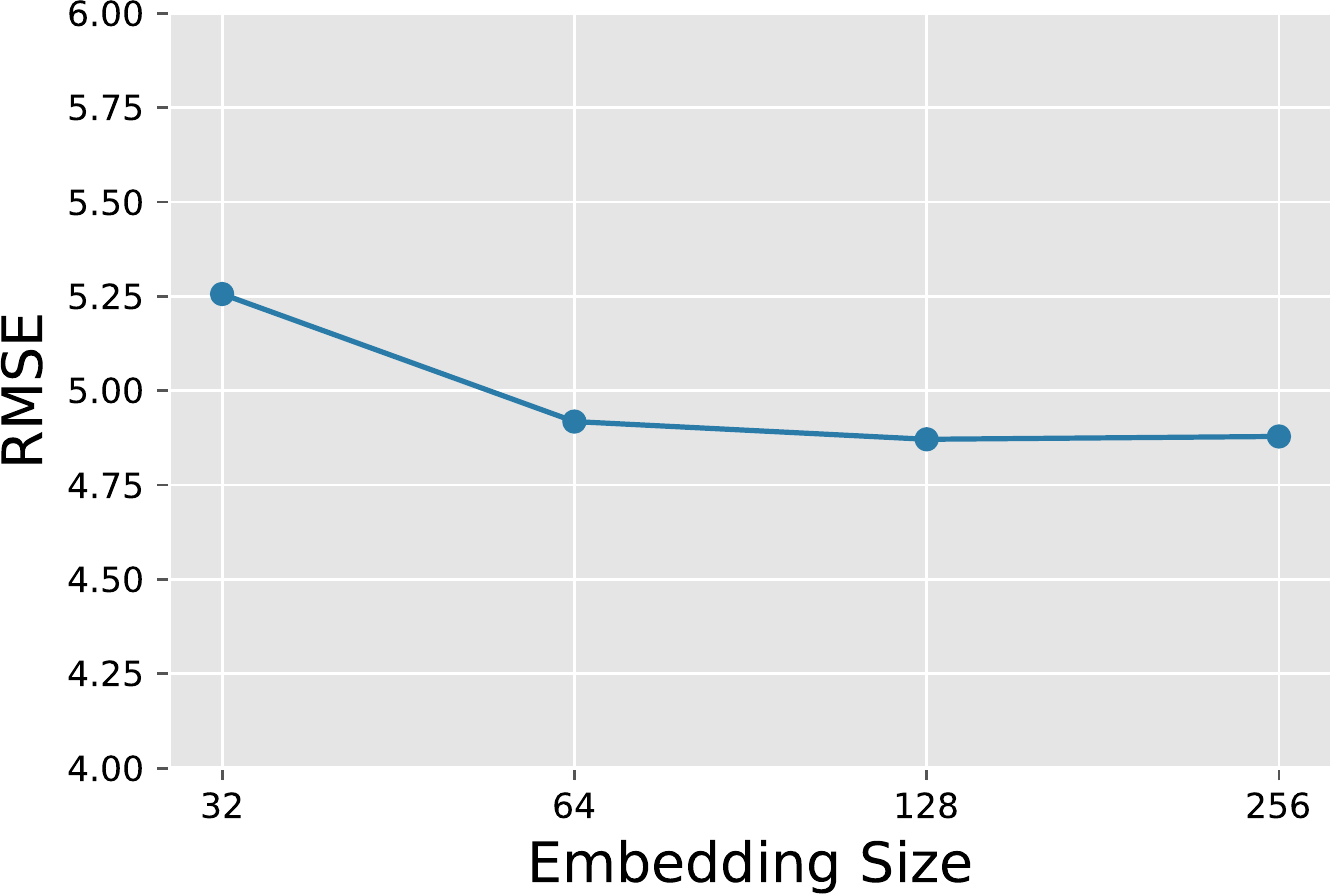}\\[-0.25in]
            \label{fig:hyper-embsize}
        \end{minipage}
    }
    %\par
    \subfloat[Effect of multitask weights.]{
        \begin{minipage}[t]{0.33\linewidth}
            \centering
            \includegraphics[width=1.0\linewidth]{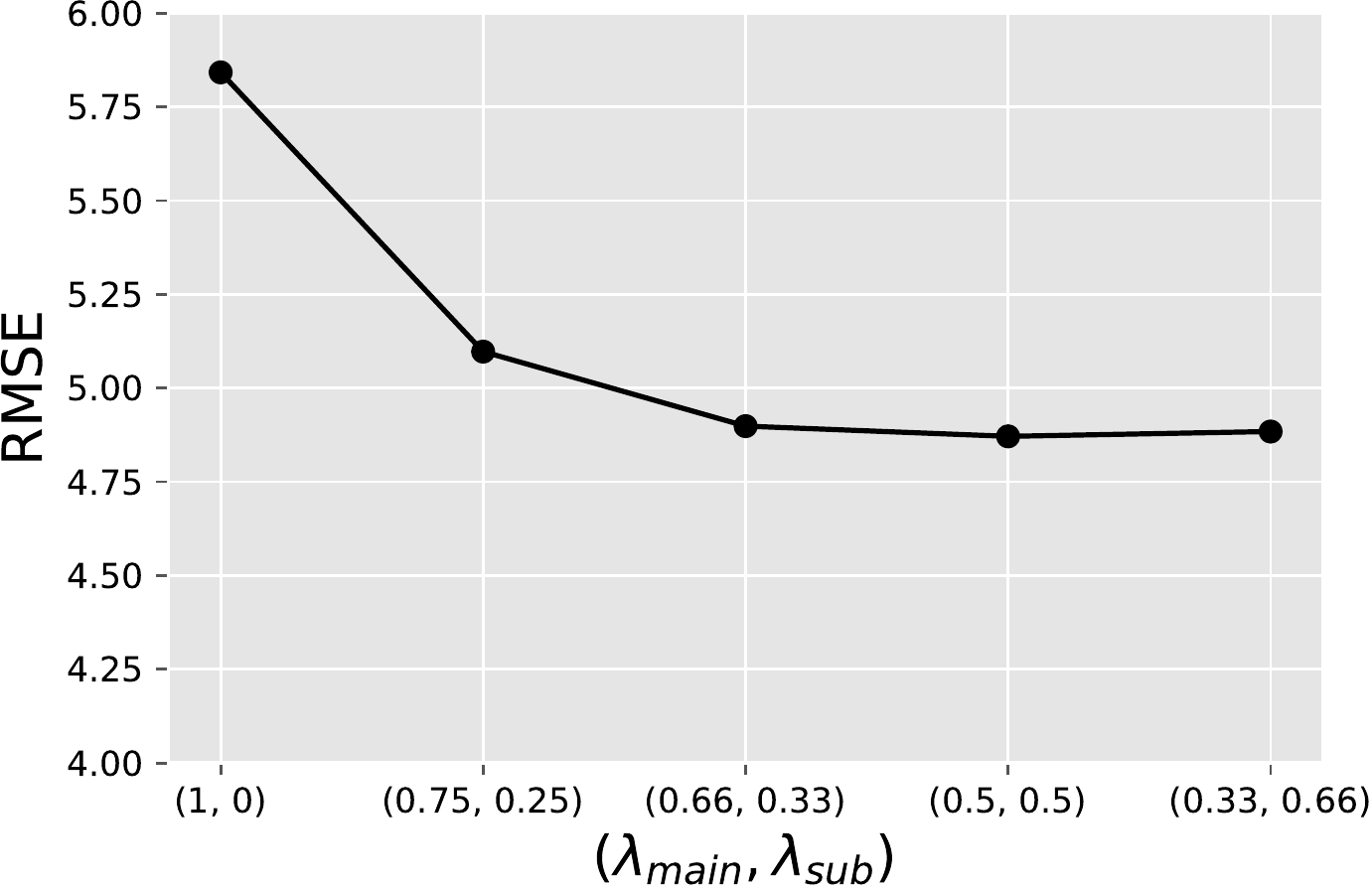}\\[-0.25in]
            \label{fig:hyper-multitask}
        \end{minipage}
    }
    % \vspace{-0.15in}
    \caption{Results of different hyperparameter settings.}
    \label{fig:param sensitivity}
\end{figure*}

%% file: performance.tex
% Table generated by Excel2LaTeX from sheet 'performance'
\begin{table}[htbp]
  \centering
  \begin{threeparttable}
  \caption{Performance on Test Set}
    \begin{tabular}{cccc}
    \toprule
    \textit{\textbf{Model}} & \textbf{RMSE} & \textbf{MAE} & \textbf{CPC}\tnote{*} \\
    \midrule
    \midrule
    GM-P  & 7.035 & 2.236 & 0.589 \\
    \midrule
    GM-E  & 6.944 & 2.179 & 0.602 \\
    \midrule
    RF    & 6.273 & 2.436 & 0.638 \\
    \midrule
    GBRT  & 5.454 & 1.974 & 0.707 \\
    \midrule
    Node2vec    & 5.455 & 1.994 & 0.704 \\
    \midrule
    GMEL-noMul & 5.356 & 1.910 & 0.716 \\
    \midrule
    GMEL-noSep & 4.982 & 1.772 & 0.737 \\
    \midrule
    \textbf{GMEL~(ours)} & \textbf{4.887} & \textbf{1.747} & \textbf{0.741} \\
    \bottomrule
    \end{tabular}%
    \begin{tablenotes}\footnotesize
        \item[*] Higher is better.
    \end{tablenotes}
  \label{tab:performance}%
  \end{threeparttable}
\end{table}%

% % Table generated by Excel2LaTeX from sheet 'performance'
% \begin{table}[htbp]
%   \centering
%   \begin{threeparttable}
%   \caption{Performance on Test Set}
%     \begin{tabular}{cccc}
%     \toprule
%     \textit{\textbf{Model}} & \textbf{RMSE} & \textbf{MAE} & \textbf{\makecell{CPC\\(the larger\\the better)}} \\
%     \midrule
%     \midrule
%     GM-P  & 7.035 & 2.236 & 0.589 \\
%     \midrule
%     GM-E  & 6.944 & 2.179 & 0.602 \\
%     \midrule
%     RF    & 6.273 & 2.436 & 0.638 \\
%     \midrule
%     GBRT  & 5.454 & 1.974 & 0.707 \\
%     \midrule
%     GMEL-noMul & 5.356 & 1.910 & 0.716 \\
%     \midrule
%     GMEL-noSep & 4.982 & 1.772 & 0.737 \\
%     \midrule
%     \textbf{GMEL} & \textbf{4.887} & \textbf{1.747} & \textbf{0.741} \\
%     \bottomrule
%     \end{tabular}%
%   \label{tab:performance}%
%   \end{threeparttable}
% \end{table}%

%% file: spatial_errors.tex
\begin{figure}[hbtp]
    \centering
    
    \subfloat{
        \begin{minipage}[t]{0.5\linewidth}
            % \centering
            \includegraphics[width=\linewidth]{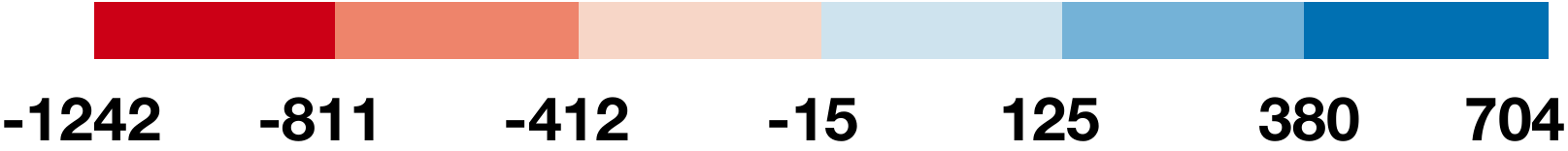}
        \end{minipage}
    }\par
    % \vspace{-1\baselineskip}
    
    \setcounter{subfigure}{0}
    \subfloat[Residuals of GBRT]{
        \begin{minipage}[t]{0.5\linewidth}
            \centering
            \includegraphics[width=\linewidth]{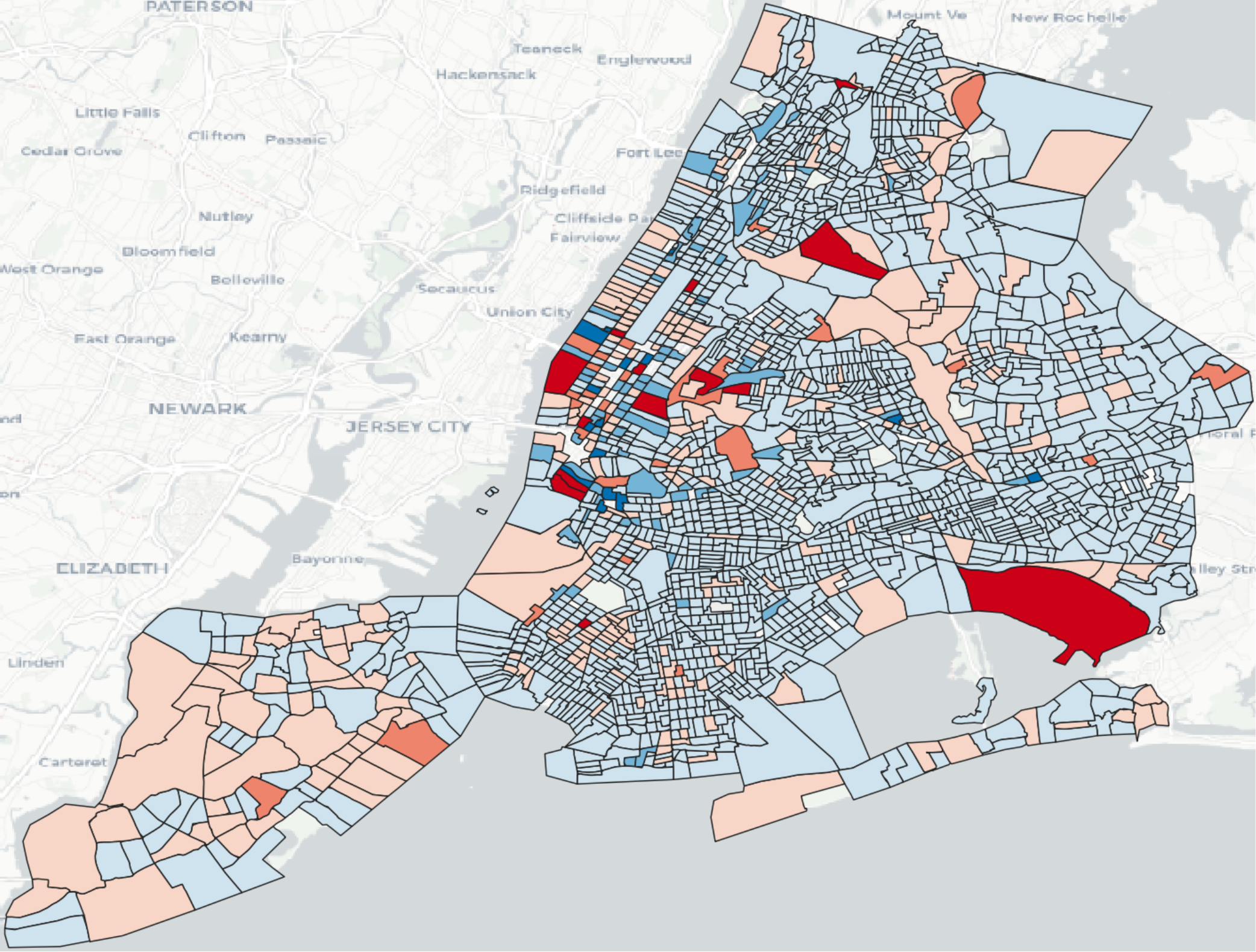}\\[-0.25in]
            \label{fig:spatial error gbrt}
        \end{minipage}
    }
    \subfloat[Residuals of GMEL]{
        \begin{minipage}[t]{0.5\linewidth}
            \centering
            \includegraphics[width=\linewidth]{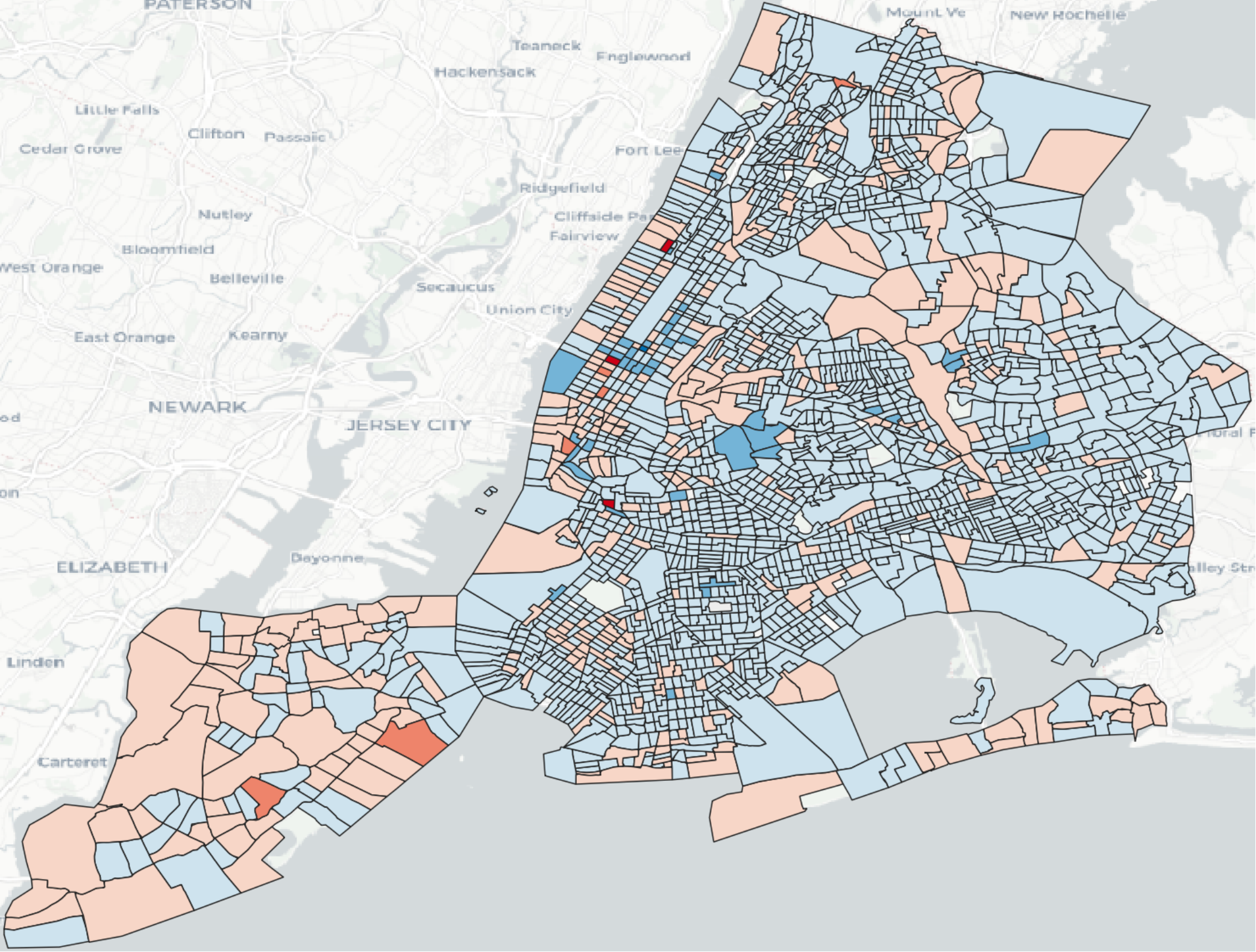}\\[-0.25in]
            \label{fig:spatial error ours}
        \end{minipage}
    }
    % \vspace{-0.1in}
    \caption{Spatial distribution of residuals. Red indicates underestimation and blue indicates overestimation. Light blue census tracts indicate the best predictions.}
    \label{fig:spatial error}
\end{figure}

%% file: feature_saliency.tex
\begin{figure*}[t!]
    \centering
    \subfloat[Edge-level flow.]{
        \begin{minipage}[t]{0.33\linewidth}
            \centering
            \includegraphics[height=80px]{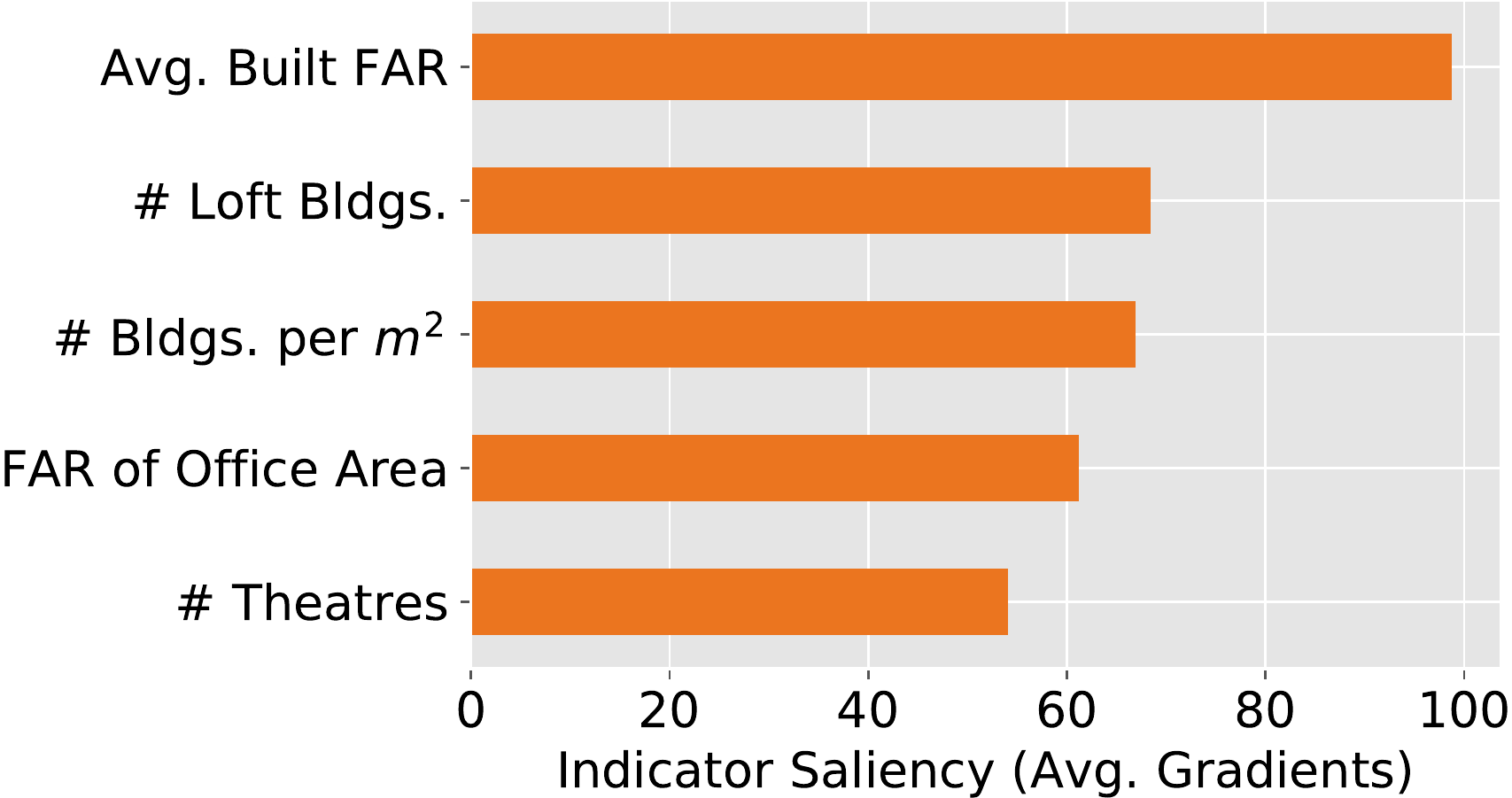}\\[-0.25in]
            \label{fig:edgesaliency}
        \end{minipage}
    }
    \subfloat[In-flow.]{
        \begin{minipage}[t]{0.33\linewidth}
            \centering
            \includegraphics[height=80px]{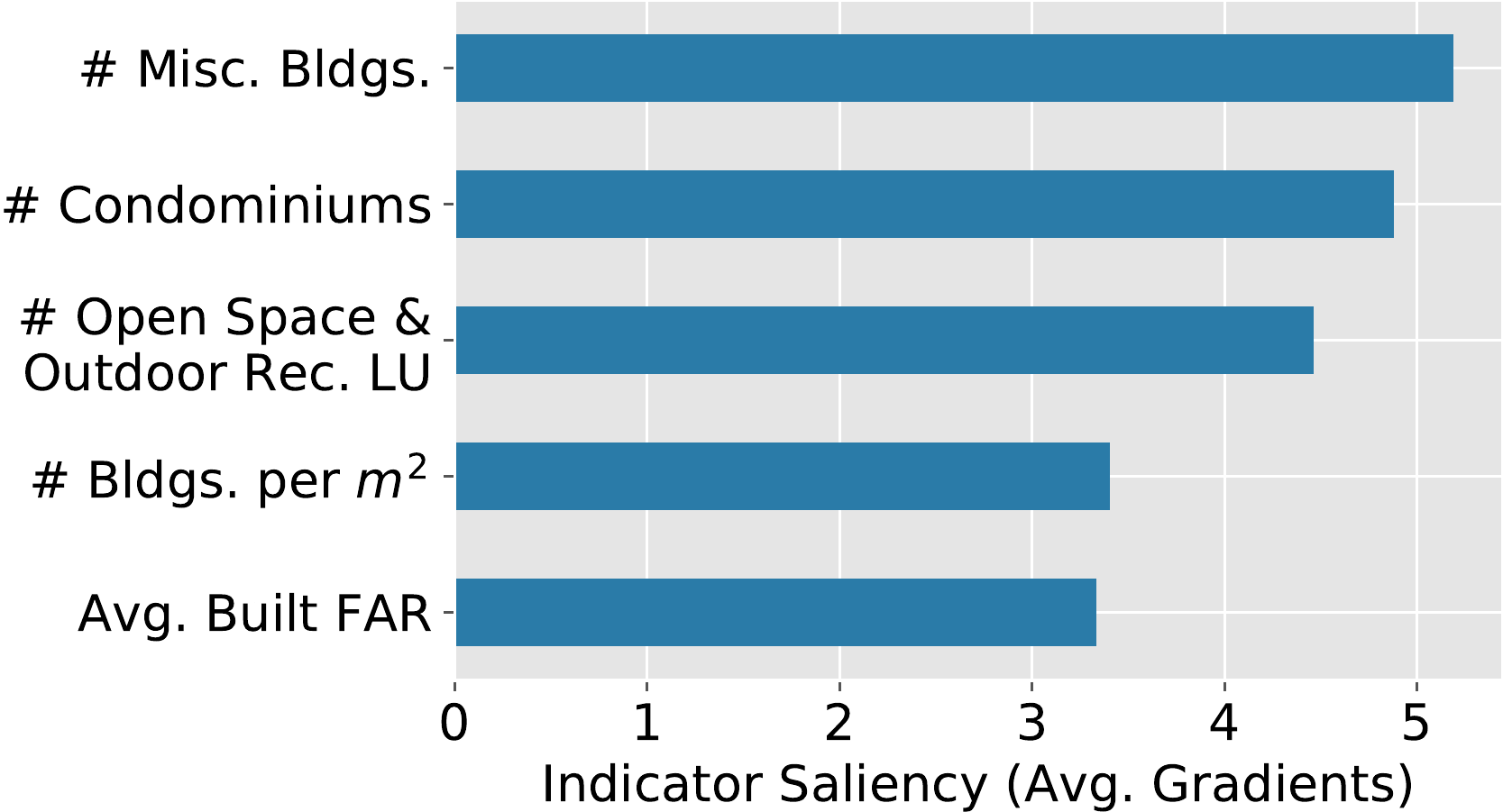}\\[-0.25in]
            \label{fig:insaliency}
        \end{minipage}
    }
    %\par
    \subfloat[Out-flow.]{
        \begin{minipage}[t]{0.33\linewidth}
            \centering
            \includegraphics[height=80px]{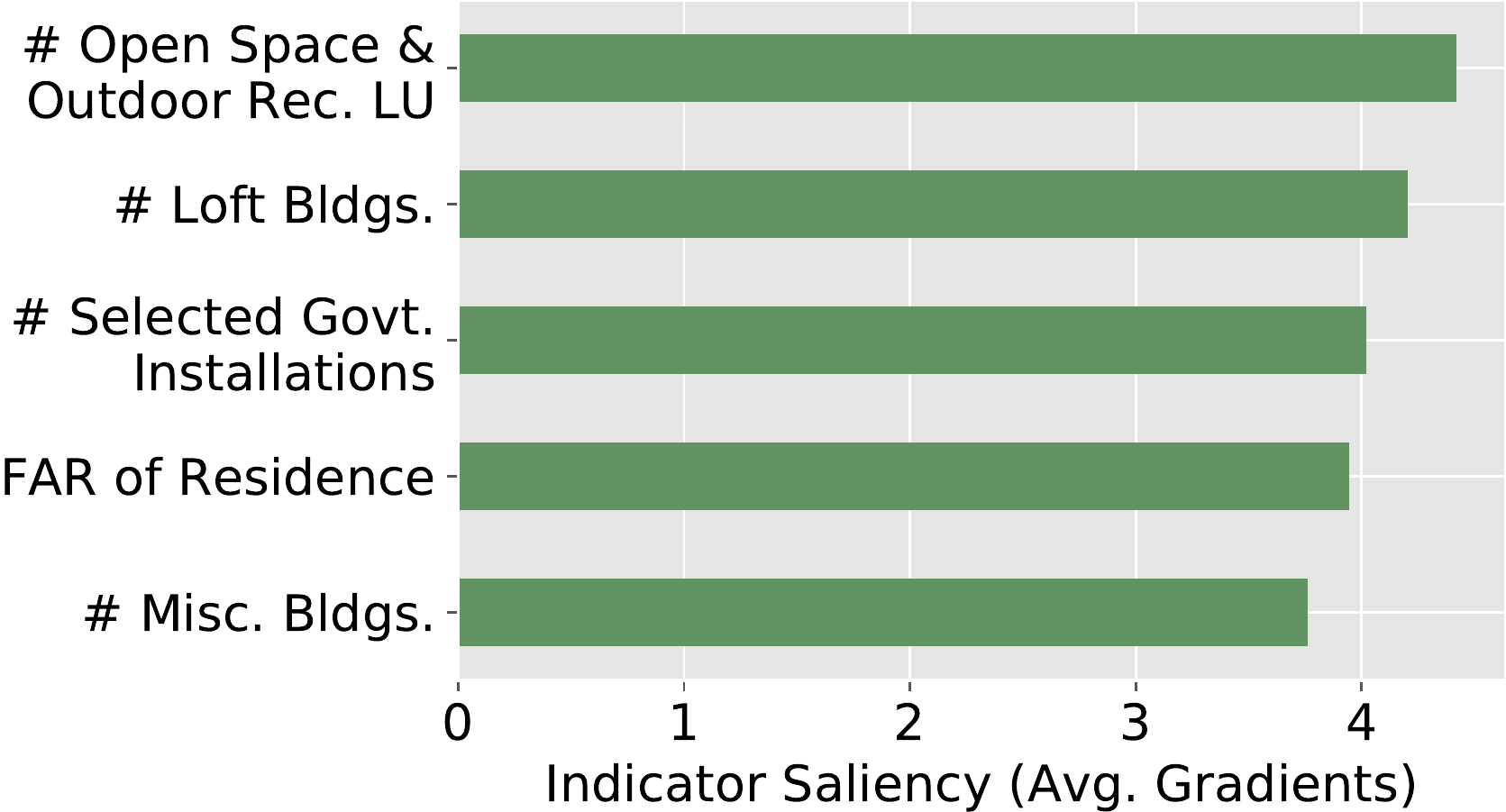}\\[-0.25in]
            \label{fig:outsaliency}
        \end{minipage}
    }
    % \vspace{-0.15in}
    \caption{Top-5 salient urban indicators.}
    \label{fig:saliency}
\end{figure*}

%% file: 06conclusion.tex
\section{Conclusion}

In this paper, we study the problem of predicting commuting flow using only the information of infrastructure and land use, a fundamental problem in urban planning and public policy development. Different from conventional gravity model and recently proposed machine learning methods, we propose the use of geographic contextual information for commuting flow prediction. As such, an end-to-end embedding learning framework based on graph attention network is proposed to learn geo-contextual embeddings of the geographic units. The learned embeddings are then fed to a gradient boosting machine to make predictions. We conduct extensive experiments on real-world datasets from New~York~City. The results show that introducing geographic contextual information can greatly improve the accuracy of prediction and our model outperforms all baseline methods including the state of the art.

%% file: 07acknowledgement.tex
\section{ Acknowledgements}

This  work  was  supported  in  part  by:  the  Moore-Sloan  Data Science  Environment  at  NYU;  NASA;  DOE;  National Science Foundation awards CNS-1229185, CCF-1533564, CNS-1544753,  CNS-1730396, MRI-1229185; State Key Program of National Natural Science Foundation of China under grant No. 51838002; National Natural Science Foundation of China under Grant No. 51578128; National Science and Technology Major Project of China under Grant 2016ZX03001022-002; Program of China Scholarships Council No. 201806090079.
C.~T.~Silva is partially supported by the DARPA D3M program. Any opinions, findings, and conclusions or recommendations expressed in this material are those of the authors and do not necessarily reflect the views of DARPA.
We also gratefully acknowledge the support of NVIDIA Corporation with the donation of some of the GPUs used in this research.

%% file: AAAI-LiuZ.9223.bbl
\begin{thebibliography}{}

\bibitem[\protect\citeauthoryear{Albeverio \bgroup et al\mbox.\egroup
  }{2007}]{albeverio_dynamics_2007}
Albeverio, S.; Andrey, D.; Giordano, P.; and Vancheri, A.
\newblock 2007.
\newblock {\em The dynamics of complex urban systems: {An} interdisciplinary
  approach}.
\newblock Springer.

\bibitem[\protect\citeauthoryear{Austin}{2017}]{doi:10.1002/oa.2575}
Austin, A.~E.
\newblock 2017.
\newblock The cost of a commute: A multidisciplinary approach to osteoarthritis
  in new kingdom egypt.
\newblock {\em International Journal of Osteoarchaeology} 27(4):537--550.

\bibitem[\protect\citeauthoryear{Bram and McKay}{2005}]{nyccommute}
Bram, J., and McKay, A.
\newblock 2005.
\newblock Evolution of commuting patterns in the {New York City} metro area.
\newblock {\em Current Issues in Economics and Finance}.

\bibitem[\protect\citeauthoryear{Caruana}{1997}]{caruana_multitask_1997}
Caruana, R.
\newblock 1997.
\newblock Multitask learning.
\newblock {\em Machine Learning} 28(1):41--75.

\bibitem[\protect\citeauthoryear{Friedman, Hastie, and
  Tibshirani}{2001}]{friedman_elements_2001}
Friedman, J.; Hastie, T.; and Tibshirani, R.
\newblock 2001.
\newblock {\em The elements of statistical learning}, volume~1.
\newblock Springer Series in Statistics, New York, NY, USA.

\bibitem[\protect\citeauthoryear{Grover and
  Leskovec}{2016}]{grover_node2vec:_2016}
Grover, A., and Leskovec, J.
\newblock 2016.
\newblock Node2vec: {Scalable} {Feature} {Learning} for {Networks}.
\newblock In {\em Proceedings of the 22Nd {ACM} {SIGKDD} {International}
  {Conference} on {Knowledge} {Discovery} and {Data} {Mining}}, {KDD} '16,
  855--864.
\newblock New York, NY, USA: ACM.

\bibitem[\protect\citeauthoryear{Hamilton, Ying, and
  Leskovec}{2017}]{hamilton_inductive_2017}
Hamilton, W.; Ying, Z.; and Leskovec, J.
\newblock 2017.
\newblock Inductive representation learning on large graphs.
\newblock In {\em Advances in Neural Information Processing Systems 30}. Curran
  Associates, Inc.
\newblock  1024--1034.

\bibitem[\protect\citeauthoryear{Iwata and Hitoshi}{2019}]{iwata}
Iwata, T., and Hitoshi, S.
\newblock 2019.
\newblock Neural collective graphical models for estimating spatio-temporal
  population flow from aggregated data.
\newblock {\em Proceedings of the AAAI Conference on Artificial Intelligence}
  33:3935--3942.

\bibitem[\protect\citeauthoryear{Ke \bgroup et al\mbox.\egroup
  }{2019}]{ke_deepgbm:_2019}
Ke, G.; Xu, Z.; Zhang, J.; Bian, J.; and Liu, T.-Y.
\newblock 2019.
\newblock {DeepGBM}: A deep learning framework distilled by {GBDT} for online
  prediction tasks.
\newblock In {\em Proceedings of the 25th {ACM} {SIGKDD} {International}
  {Conference} on {Knowledge} {Discovery} \& {Data} {Mining}}, {KDD} '19,
  384--394.
\newblock New York, NY, USA: ACM.

\bibitem[\protect\citeauthoryear{Lenormand, Bassolas, and
  Ramasco}{2016}]{lenormand_systematic_2016}
Lenormand, M.; Bassolas, A.; and Ramasco, J.~J.
\newblock 2016.
\newblock Systematic comparison of trip distribution laws and models.
\newblock {\em Journal of Transport Geography} 51:158--169.

\bibitem[\protect\citeauthoryear{Luxen and Vetter}{2011}]{luxen_real-time_2011}
Luxen, D., and Vetter, C.
\newblock 2011.
\newblock Real-time routing with openstreetmap data.
\newblock In {\em Proceedings of the 19th {ACM} {SIGSPATIAL} {International}
  {Conference} on {Advances} in {Geographic} {Information} {Systems}}, {GIS}
  '11,  513--516.
\newblock New York, NY, USA: ACM.

\bibitem[\protect\citeauthoryear{Monge}{1781}]{monge_memoire_1781}
Monge, G.
\newblock 1781.
\newblock Mémoire sur la théorie des déblais et des remblais.
\newblock {\em Histoire de l'Académie Royale des Sciences de Paris}.

\bibitem[\protect\citeauthoryear{{NYC DCP}}{2015}]{pluto}
{NYC DCP}.
\newblock 2015.
\newblock Pluto and mappluto.
\newblock
  \url{https://www1.nyc.gov/site/planning/data-maps/open-data/dwn-pluto-mappluto.page}.

\bibitem[\protect\citeauthoryear{{Office of the State Deputy Comptroller for
  NYC}}{2019}]{employment}
{Office of the State Deputy Comptroller for NYC}.
\newblock 2019.
\newblock {New York City} employment trends.

\bibitem[\protect\citeauthoryear{Pan \bgroup et al\mbox.\egroup
  }{2019}]{pan_urban_2019}
Pan, Z.; Liang, Y.; Wang, W.; Yu, Y.; Zheng, Y.; and Zhang, J.
\newblock 2019.
\newblock Urban traffic prediction from spatio-temporal data using deep meta
  learning.
\newblock In {\em Proceedings of the 25th {ACM} {SIGKDD} {International}
  {Conference} on {Knowledge} {Discovery} \& {Data} {Mining}}, {KDD} '19,
  1720--1730.
\newblock New York, NY, USA: ACM.

\bibitem[\protect\citeauthoryear{Paszke \bgroup et al\mbox.\egroup
  }{2017}]{paszke2017automatic}
Paszke, A.; Gross, S.; Chintala, S.; Chanan, G.; Yang, E.; DeVito, Z.; Lin, Z.;
  Desmaison, A.; Antiga, L.; and Lerer, A.
\newblock 2017.
\newblock Automatic differentiation in pytorch.
\newblock In {\em NeurIPS Workshop}.

\bibitem[\protect\citeauthoryear{Pourebrahim \bgroup et al\mbox.\egroup
  }{2019}]{pourebrahim_trip_2019}
Pourebrahim, N.; Sultana, S.; Niakanlahiji, A.; and Thill, J.-C.
\newblock 2019.
\newblock Trip distribution modeling with {Twitter} data.
\newblock {\em Computers, Environment and Urban Systems} 77:101354.

\bibitem[\protect\citeauthoryear{Robinson and
  Dilkina}{2018}]{robinson_machine_2018}
Robinson, C., and Dilkina, B.
\newblock 2018.
\newblock A machine learning approach to modeling human migration.
\newblock In {\em Proceedings of the 1st ACM SIGCAS Conference on Computing and
  Sustainable Societies}, COMPASS '18,  30:1--30:8.
\newblock New York, NY, USA: ACM.

\bibitem[\protect\citeauthoryear{Rodrigue, Comtois, and
  Slack}{2016}]{rodrigue_geography_2016}
Rodrigue, J.-P.; Comtois, C.; and Slack, B.
\newblock 2016.
\newblock {\em The geography of transport systems}.
\newblock Routledge.

\bibitem[\protect\citeauthoryear{Simini \bgroup et al\mbox.\egroup
  }{2012}]{simini_universal_2012}
Simini, F.; González, M.~C.; Maritan, A.; and Barabási, A.-L.
\newblock 2012.
\newblock A universal model for mobility and migration patterns.
\newblock {\em Nature} 484(7392):96--100.

\bibitem[\protect\citeauthoryear{Simonyan, Vedaldi, and
  Zisserman}{2013}]{simonyan2013deep}
Simonyan, K.; Vedaldi, A.; and Zisserman, A.
\newblock 2013.
\newblock Deep inside convolutional networks: Visualising image classification
  models and saliency maps.
\newblock {\em arXiv preprint arXiv:1312.6034}.

\bibitem[\protect\citeauthoryear{Spadon \bgroup et al\mbox.\egroup
  }{2019}]{spadon_reconstructing_2019}
Spadon, G.; Carvalho, A. C. P. L. F.~d.; Rodrigues-Jr, J.~F.; and Alves, L.
  G.~A.
\newblock 2019.
\newblock Reconstructing commuters network using machine learning and urban
  indicators.
\newblock {\em Scientific Reports} 9(1):1--13.

\bibitem[\protect\citeauthoryear{Tobler}{1970}]{tobler_computer_1970}
Tobler, W.
\newblock 1970.
\newblock A computer movie simulating urban growth in the detroit region.
\newblock {\em Economic Geography} 46(2):234--240.

\bibitem[\protect\citeauthoryear{{US Census Bureau}}{2015}]{lehd}
{US Census Bureau}.
\newblock 2015.
\newblock Longitudinal employer-household dynamics.
\newblock \url{https://lehd.ces.census.gov/data/}.

\bibitem[\protect\citeauthoryear{Veli{\v{c}}kovi{\'{c}} \bgroup et
  al\mbox.\egroup }{2018}]{velickovic_graph_2017}
Veli{\v{c}}kovi{\'{c}}, P.; Cucurull, G.; Casanova, A.; Romero, A.; Li{\`{o}},
  P.; and Bengio, Y.
\newblock 2018.
\newblock {Graph Attention Networks}.
\newblock {\em International Conference on Learning Representations}.

\bibitem[\protect\citeauthoryear{Wang and
  Li}{2017}]{Wang:2017:RRL:3132847.3133006}
Wang, H., and Li, Z.
\newblock 2017.
\newblock Region representation learning via mobility flow.
\newblock In {\em Proceedings of the 2017 ACM on Conference on Information and
  Knowledge Management}, CIKM '17,  237--246.
\newblock New York, NY, USA: ACM.

\bibitem[\protect\citeauthoryear{Wang \bgroup et al\mbox.\egroup
  }{2019a}]{wang2019dgl}
Wang, M.; Yu, L.; Zheng, D.; Gan, Q.; Gai, Y.; Ye, Z.; Li, M.; Zhou, J.; Huang,
  Q.; Ma, C.; Huang, Z.; Guo, Q.; Zhang, H.; Lin, H.; Zhao, J.; Li, J.; Smola,
  A.~J.; and Zhang, Z.
\newblock 2019a.
\newblock Deep graph library: Towards efficient and scalable deep learning on
  graphs.
\newblock {\em ICLR Workshop on Representation Learning on Graphs and
  Manifolds}.

\bibitem[\protect\citeauthoryear{Wang \bgroup et al\mbox.\egroup
  }{2019b}]{wang_origin-destination_2019}
Wang, Y.; Yin, H.; Chen, H.; Wo, T.; Xu, J.; and Zheng, K.
\newblock 2019b.
\newblock Origin-destination matrix prediction via graph convolution: A new
  perspective of passenger demand modeling.
\newblock In {\em Proceedings of the 25th {ACM} {SIGKDD} {International}
  {Conference} on {Knowledge} {Discovery} \& {Data} {Mining}}, {KDD} '19,
  1227--1235.
\newblock New York, NY, USA: ACM.

\bibitem[\protect\citeauthoryear{Xiong \bgroup et al\mbox.\egroup
  }{}]{xiong_dynamic_nodate}
Xiong, X.; Ozbay, K.; Jin, L.; and Feng, C.
\newblock Dynamic origin-destination matrix prediction with line graph neural
  networks and {Kalman} filter.
\newblock ~23.

\bibitem[\protect\citeauthoryear{Yang \bgroup et al\mbox.\egroup
  }{2014}]{yang2014limits}
Yang, Y.; Herrera, C.; Eagle, N.; and Gonz{\'a}lez, M.~C.
\newblock 2014.
\newblock Limits of predictability in commuting flows in the absence of data
  for calibration.
\newblock {\em Scientific reports} 4:5662.

\bibitem[\protect\citeauthoryear{Zhang \bgroup et al\mbox.\egroup
  }{2019}]{zhang_flow_2019}
Zhang, J.; Zheng, Y.; Sun, J.; and Qi, D.
\newblock 2019.
\newblock Flow prediction in spatio-temporal networks based on multitask deep
  learning.
\newblock {\em IEEE Transactions on Knowledge and Data Engineering}  1--1.

\bibitem[\protect\citeauthoryear{Zipf}{1946}]{zipf_p1_1946}
Zipf, G.~K.
\newblock 1946.
\newblock The {P}1 {P}2/{D} hypothesis: On the intercity movement of persons.
\newblock {\em American Sociological Review} 11(6):677--686.

\end{thebibliography}
